\documentclass[a4paper,12pt]{article}
\usepackage{amssymb}
\usepackage{amsmath}
\usepackage{amsfonts}
\usepackage{mathtools}
\usepackage{slashed}
\usepackage{color}
\usepackage{indentfirst}
\usepackage{graphicx}
\usepackage{csquotes} 
\usepackage{caption}
\usepackage{cite}
\usepackage{here}
\usepackage{comment}
\usepackage{listings}
\usepackage[italicdiff]{physics}
\makeatletter
\newcommand*\rel@kern[1]{\kern#1\dimexpr\macc@kerna}
\newcommand*\widebar[1]{%
  \begingroup
  \def\mathaccent##1##2{%
    \rel@kern{0.8}%
    \overline{\rel@kern{-0.8}\macc@nucleus\rel@kern{0.2}}%
    \rel@kern{-0.2}%
  }%
  \macc@depth\@ne
  \let\math@bgroup\@empty \let\math@egroup\macc@set@skewchar
  \mathsurround\z@ \frozen@everymath{\mathgroup\macc@group\relax}%
  \macc@set@skewchar\relax
  \let\mathaccentV\macc@nested@a
  \macc@nested@a\relax111{#1}%
  \endgroup
}
\makeatother

\renewcommand{\thefootnote}{\fnsymbol{footnote}}

\begin{document}

\title{
\begin{flushright}
\begin{minipage}{0.2\linewidth}
\normalsize
EPHOU-21-004\\
KEK-TH-2312 \\
WU-HEP-21-01\\*[50pt]
\end{minipage}
\end{flushright}
{\Large \bf 
Hierarchical structure of physical Yukawa couplings
from matter field K\"ahler metric 
\\*[20pt]}}

\author{Keiya Ishiguro$^{a}$\footnote{
E-mail address: keyspire@ruri.waseda.jp
},\,
Tatsuo Kobayashi$^{b}$\footnote{
E-mail address: kobayashi@particle.sci.hokudai.ac.jp
}
\ and\
Hajime Otsuka$^{c}$\footnote{
E-mail address: hotsuka@post.kek.jp
}\\*[20pt]
$^a${\it \normalsize 
Department of Physics, Waseda University, Tokyo 169-8555, Japan} \\
$^b${\it \normalsize 
Department of Physics, Hokkaido University, Sapporo 060-0810, Japan} \\
$^c${\it \normalsize 
KEK Theory Center, Institute of Particle and Nuclear Studies, KEK,}\\
{\it \normalsize 1-1 Oho, Tsukuba, Ibaraki 305-0801, Japan}}

\maketitle

\date{
\centerline{\small \bf Abstract}
\begin{minipage}{0.9\linewidth}
\medskip 
\medskip 
\small
We study the impacts of matter field K\"ahler metric on physical 
Yukawa couplings in string compactifications. 
Since the K\"ahler metric is non-trivial in general, 
the kinetic mixing of matter fields opens a new avenue for realizing a  hierarchical structure of physical Yukawa couplings, even when holomorphic 
Yukawa couplings have the trivial structure. 
The hierarchical Yukawa couplings are demonstrated 
by couplings of pure untwisted modes on toroidal orbifolds and their resolutions 
in the context of heterotic 
string theory with standard embedding. 
Also, we study the hierarchical couplings among untwisted and twisted modes 
on resolved orbifolds.
\end{minipage}
}

\renewcommand{\thefootnote}{\arabic{footnote}}
\setcounter{footnote}{0}
%\vspace{35pt}
\thispagestyle{empty}
\clearpage
\addtocounter{page}{-1}
%\newpage
%\setcounter{tocdepth}{3}

\tableofcontents

%%%%%%%%%%%%%%%%%%%%%%%%%%%%%%%%%%%%%%%%%%%%%%%%%%%%%%%%%%%%
%%%%%%%%%%%%%%%%%%%%%%%%%%%%%%%%%%%%%%%%%%%%%%%%%%%%%%%%%%%%
\section{Introduction}
\label{sec:Intro}
%%%%%%%%%%%%%%%%%%%%%%%%%%%%%%%%%%%%%%%%%%%%%%%%%%%%%%%%%%%%
%%%%%%%%%%%%%%%%%%%%%%%%%%%%%%%%%%%%%%%%%%%%%%%%%%%%%%%%%%%%

The origin of flavor structure of quarks and leptons, in particular 
the hierarchical structure of their masses is 
one of unsolved issues in the Standard Model (SM) of particle physics. 
The Yukawa couplings of quarks and leptons are free parameters such that 
their masses and mixing angles are fitted with experimental data. 
There is no guiding principle to choose the Yukawa couplings in the quark sector, 
which have hierarchical ratios of ${\cal O}(10^5)$.
Several mechanisms are proposed to explain the structure of Yukawa couplings, 
as represented by the localization of matter wavefunctions in the extra-dimensional space \cite{ArkaniHamed:1999dc,Kaplan:2000av} 
and a proper charge assignment of quarks/leptons under the continuous or 
discrete flavor symmetries \cite{Froggatt:1978nt,Ishimori:2010au}. 
These mechanisms can be engineered in the ultra-violet completion of 
the SM, such as its supersymmetric extension and the string theory.

In the four-dimensional (4D) ${\cal N}=1$ supersymmetric theory,
physical Yukawa couplings $Y_{\hat{a}\hat{b}\hat{c}}$ are 
obtained from holomorphic Yukawa couplings $\kappa_{abc}$
in the canonical basis, where a kinetic term of matter fields is canonical, i.e., 
the diagonal K\"ahler metric. 
Note that the K\"ahler metric is in a non-diagonal, but non-trivial form in a 
generic supergravity theory derived from the superstring theory. 
There are two possibilities to realize the hierarchical structure of physical Yukawa couplings: (i) the flavor structure of holomorphic Yukawa 
couplings controlled by the continuous/discrete flavor symmetries, 
(ii) a non-trivial structure of matter field K\"ahler metric which 
would be connected with the geometry of extra-dimensional space.

In the traditional approach (i), 
holomorphic Yukawa couplings are well controlled by stringy selection 
rules, gauge and  extra-dimensional geometrical symmetries. 
It was known that the holomorphic Yukawa couplings of twisted modes 
on heterotic orbifold models are exponentially suppressed with respect 
to moduli fields \cite{Hamidi:1986vh,Dixon:1986qv,Burwick:1990tu}. 
Hence, one can realize the hierarchical structure of physical Yukawa 
couplings of these states due to the fact that they are localized at fixed 
points.
However, the coupling selection rules among twisted modes are 
very tight \cite{Kobayashi:1991rp,Casas:1991ac}.
It is not straightforward to realize realistic mass matrices 
for three generations of quarks and leptons.
(See for attempts to realize quark and lepton mass matrices 
by twisted modes, Refs.~\cite{Ko:2004ic,Ko:2005sh}.)

At any rate,  it would be difficult to have the hierarchical structure for other 
states such as untwisted modes propagating the bulk. 
It indicates that smooth Calabi-Yau (CY) compactifications will not lead to the 
hierarchical structure of holomorphic Yukawa couplings for matter zero-modes. 
For instance, in the context of heterotic string theory with standard 
embedding, 
the holomorphic Yukawa couplings are of ${\cal O}(1)$  in the large volume/complex structure regime of CY threefolds, 
determined by the third derivative of the prepotential. 
It results in the trivial structure of the holomorphic Yukawa couplings 
which are also observed in the untwisted modes on a singular limit of CY threefolds such as 
toroidal orbifolds. 
Hence, the understanding of matter field K\"ahler metric would shed new light on the structure of physical Yukawa couplings. 

In this paper, we study whether the matter field K\"ahler 
metric provides the hierarchical structure of physical Yukawa 
couplings. 
In contrast to the holomorphic Yukawa couplings, 
the matter field K\"ahler metric would not be restricted by symmetry 
arguments. 
Hence, in the approach (ii), we have to calculate zero-mode wavefunctions 
themselves living on the extra-dimensional space to understand the structure of matter field K\"ahler metric.\footnote{Explicit calculations of zero-mode wavefunctions were performed on 
e.g., toroidal backgrounds with line bundles \cite{Cremades:2004wa,Abe:2008fi,Kobayashi:2019fma} and local 4-cycle on the conifold regime of CY threefold \cite{Abe:2015bxa} in Type IIB string theory 
and smooth CY threefolds with large internal gauge fluxes \cite{Blesneag:2018ygh} in the context of heterotic string theory.} 
To overcome this problem, 
we focus on the heterotic string theory with standard embedding, 
where matter fields are in one-to-one correspondence with moduli fields on 
CY threefolds including toroidal orbifolds in the singular limit. 
In this context, the matter field K\"ahler metric as well as the holomorphic Yukawa couplings 
are explicitly calculated by means of conformal field theory as well as the special geometry \cite{Dixon:1989fj}. 
To study the impacts of matter field K\"ahler metric on physical Yukawa couplings, 
we deal with a simplified CY threefold and toroidal orbifolds preserving the supersymmetry. 
On toroidal backgrounds, the moduli fields are categorized by untwisted and twisted sectors, 
where twisted modes are analyzed through resolutions of toroidal orbifolds, capturing the 
feature of CY threefolds. 
We find that the kinetic mixings on the moduli 
space of untwisted and twisted modes 
are inevitable for a realization of hierarchical Yukawa couplings. 
Remarkably, ${\cal O}(1)$ values of the moduli fields provide the hierarchical structure of 
physical Yukawa couplings, demonstrated on specific toroidal orbifolds 
with and without resolutions having the non-diagonal matter field K\"ahler metric.

The remaining part of this paper is organized as follows. 
In Section \ref{sec:2}, we briefly review an effective action of matter fields, focusing on 
Yukawa couplings for the fundamental and anti-fundamental representations of $E_6$ gauge group. 
The off-diagonal entries in the matter field K\"ahler metric play an important role of distinguishing 
the holomorphic Yukawa couplings and the physical one as illustrated on toroidal orbifolds in detail. 
In Section \ref{sec:3}, the hierarchical physical Yukawa couplings are demonstrated on 
$T^6/\mathbb{Z}_3$ orbifold, blown-up $T^6/(\mathbb{Z}_2\times \mathbb{Z}_2)$ orbifold and a 
mirror dual of $T^6/(\mathbb{Z}_3\times \mathbb{Z}_3)$ orbifold. 
Finally, Section \ref{sec:con} is devoted to the conclusion and discussion.

%%%%%%%%%%%%%%%%%%%%%%%%%%%%%%%%%%%%%%%%%%%%%%%%%%%%%%%%%%%%
%%%%%%%%%%%%%%%%%%%%%%%%%%%%%%%%%%%%%%%%%%%%%%%%%%%%%%%%%%%%
\section{Heterotic string theory with standard embedding}
\label{sec:2}
%%%%%%%%%%%%%%%%%%%%%%%%%%%%%%%%%%%%%%%%%%%%%%%%%%%%%%%%%%%%
%%%%%%%%%%%%%%%%%%%%%%%%%%%%%%%%%%%%%%%%%%%%%%%%%%%%%%%%%%%%

In Section \ref{subsec:2_1}, we briefly review the effective action of matter fields in 
the heterotic string theory on CY compactification with standard embedding. 
In particular, we focus on Yukawa couplings of matter zero-modes, 
determined by triple intersection numbers of CY threefolds. 
In Section \ref{subsec:2_2}, the matter kinetic mixings on a simplified CY threefold 
and toroidal orbifolds are examined in the light of physical Yukawa couplings.

%%%%%%%%%%%%%%%%%%%%%%%%%%%%%%%%%%%%%%%%%%%%%%%%%%%%%%%%%%%%%%%%%%%%%%%%%%%%%%%%%%%%%%%%
\subsection{Effective action}
\label{subsec:2_1}
%%%%%%%%%%%%%%%%%%%%%%%%%%%%%%%%%%%%%%%%%%%%%%%%%%%%%%%%%%%%%%%%%%%%%%%%%%%%%%%%%%%%%%%%

The effective action of matter fields is well 
developed in the context of heterotic string theory on CY threefolds with standard embedding, 
where zero-modes are characterized by a cohomology basis of CY 
threefolds.\footnote{For more detailed discussions, we refer to e.g., Refs. \cite{Green:1987mn,Polchinski:1998rr}.} 
From the identification of $SU(3)\subset E_8$ gauge bundle with a tangent bundle 
of CY threefolds ${\cal M}$, $E_8$ gauge group is broken to $E_6$ one. 
Correspondingly, the fundamental and anti-fundamental representations of $E_6$ are 
spanned by the $H^{1,1}({\cal M})$ and $H^{2,1}({\cal M})$ cohomology bases, respectively. 

We begin with the effective action of moduli fields, whose K\"ahler 
potential is described by
\begin{align}
    K &= K_{\rm ks}(T,\widebar{T}) +K_{\rm cs}(U,\widebar{U}),
\end{align}
where $K_{\rm ks}(T,\widebar{T})$ and $K_{\rm cs}(U,\widebar{U})$ denote the K\"ahler potential 
of the K\"ahler moduli $T^a$ and the complex structure moduli $U^i$, respectively. 
It is known that they are described by a complexified K\"ahler form $J$ and a holomorphic 
three-form $\Omega$ of CY threefolds:
\begin{align}
    K_{\rm ks} &= -\ln \left(\int_{\cal M} J \wedge J \wedge J \right),
        \nonumber\\    
    K_{\rm cs} &= -\ln \left(-i \int_{\cal M} \Omega \wedge \Omega \right).
    \label{eq:Kcsks}
\end{align}
In the large volume and complex structure regime of CY threefolds, 
we can expand the complexified K\"ahler form and the holomorphic three-form on the basis of 
$H^{1,1}({\cal M})$ and $H^{2,1}({\cal M})$:
\begin{align}
    J = \sum_{a=1}^{h^{1,1}} (t^a + ib^a) w_a,\quad
    \Omega =\sum_{I=0}^{h^{2,1}} \left( X^I \alpha_I - F_I \beta^I \right),
\end{align}
from which the K\"ahler moduli are defined as $T^a \equiv t^a +ib^a$ on the basis of K\"ahler form $w_a$ with $a$ running over the number of two-cycles of CY threefolds, namely the hodge number $h^{1,1}$. 
Here, $t^a$ and $b^a$ denote the volume of two-cycle and K\"ahler axions originating from Kalb-Ramond two-form $B^{(2)}$, respectively.  

On the other hand, the holomorphic three-form $\Omega$ is expanded on the 
canonical symplectic basis $\{\alpha_I, \beta^I\}$ satisfying
\begin{align}
\int_{\cal M} \alpha_I \wedge \beta^I = \delta_I^J,\quad
\int_{\cal M} \alpha_I \wedge \alpha_J =\int_{\cal M} \beta^I \wedge \beta^J = 0,
\end{align}
on which the coefficient $F_I = \partial_I F$ corresponds to the first derivative of the prepotential $F$ with 
respect to $X^I$. 
In what follows, we adopt the gauge $X^0=1$, yielding the 
complex structure moduli in the flat coordinate: $X^i=U^i$, $(i=1,2,\cdots,h^{2,1})$ with $h^{2,1}$ being the hodge number of CY threefolds. 
There the complex structure moduli space is governed by the structure of the prepotential whose functional form is known in the large complex structure regime,
\begin{align}
F(U)  = \frac{1}{3!}\kappa_{ijk}U^iU^jU^k. 
\end{align}
We now pick up the cubic term determined by the triple intersection numbers $\kappa_{ijk}$, but other next-leading terms are irrelevant to the following discussions. Furthermore, it is possible to consider the other corner of complex 
structure moduli space, but the prepotential has a model-dependent structure as shown in an example of Section \ref{subsec:orbifold3}. 
Then, the K\"ahler potential (\ref{eq:Kcsks}) is simplified as follows:
\begin{align}
    K_{\rm cs} &= - \ln \biggl[\frac{i}{6}\kappa_{ijk}(U^i-\bar{U}^i)(U^j-\bar{U}^j)(U^k-\bar{U}^k)\biggl],
        \nonumber\\
    K_{\rm ks} &= - \ln \biggl[\frac{1}{6}\kappa_{abc} (T^a+\bar{T}^a)(T^b+\bar{T}^b)(T^c+\bar{T}^c)\biggl],
\end{align}
where it is notable that loop and quantum corrections are assumed to be enough suppressed in comparison with the 
classical terms. 

We next deal with the effective action of charged matter fields $27$ and $\widebar{27}$ under the $E_6$ gauge group, 
which are denoted by $A^a$ and $A^i$, respectively. 
The matter K\"ahler potential is only known in an expansion of $A^a$ and $A^i$, 
that is, $A^a, A^i \ll 1$ \cite{Dixon:1989fj}, and the corresponding K\"ahler metrics become
\begin{align}
   K^{(27)}_{a\bar{b}} &= e^{\frac{1}{3}(K_{\rm cs}-K_{\rm ks})}(K_{\rm ks})_{a\bar{b}},
    \nonumber\\
    K^{(\widebar{27})}_{i\bar{j}} &= e^{-\frac{1}{3}(K_{\rm cs}-K_{\rm ks})}(K_{\rm cs})_{i\bar{j}}.
\end{align}
The matter field K\"ahler metric is in turn in one-to-one correspondence with the moduli K\"ahler metric. 
In addition, Yukawa interactions of charged matter fields are described by the following superpotential:
\begin{align}
     W &= \kappa_{abc}A^aA^bA^c + \kappa_{ijk}A^iA^jA^k,
\end{align}
stating that the holomorphic $27^3$ and $\widebar{27}^3$ Yukawa couplings are determined by the 
triple intersection numbers of the K\"ahler and complex structure moduli, respectively \cite{Strominger:1985it,Candelas:1987se}. 
To derive the physical Yukawa couplings, we perform the unitary transformation $L$ diagonalizing the kinetic terms for multiple $27$ matter fields,\footnote{Here, we focus on the $27$ sector, 
but it is straightforward to extend the following calculation to $\widebar{27}$ sector. 
Furthermore, we note that the hat index is introduced to canonically normalized fields in subsequent analyses.}
\begin{align}
A^a K^{(27)}_{a\bar{b}} \bar{A}^{\bar{b}} 
=e^{\frac{1}{3}(K_{\rm cs}-K_{\rm ks})} A^a  
(L^{\dagger})_{a}^{\bar{\hat{c}}}
\Lambda_{\bar{\hat{c}}\hat{c}} L^{\hat{c}}_{\bar{b}} 
\bar{A}^{\bar{b}} 
\equiv |{\cal A}^{\hat{a}}|^2,
\end{align}
where $\Lambda$ stands for an eigenvalue matrix satisfying $L KL^{\dagger} = \Lambda$. 
The canonically normalized matter fields ${\cal A}^{\hat{a}}$ are defined as 
\begin{align}
&{\cal A}^{\hat{a}} = e^{\frac{1}{6}(K_{\rm cs}-K_{\rm ks})} A^c (L^{\dagger})_c^{\hat{a}} (\sqrt{\Lambda})_{\hat{a}\bar{\hat{a}}}
=e^{\frac{1}{6}(K_{\rm cs}-K_{\rm ks})} A^c (L^{\dagger}\sqrt{\Lambda})_c^{\hat{a}},
\end{align}
and the physical Yukawa couplings result in
\begin{align}
&e^{K/2}\kappa_{def}A^dA^eA^f= e^{K_{\rm ks}}\kappa_{def} {\cal A}^{\hat{a}}{\cal A}^{\hat{b}}{\cal A}^{\hat{c}} 
   (\Lambda^{-1/2}L)_{\hat{a}}^{d}  (\Lambda^{-1/2}L)_{\hat{b}}^{e} (\Lambda^{-1/2}L)_{\hat{c}}^{f}
    \equiv Y_{\hat{a}\hat{b}\hat{c}} {\cal A}^{\hat{a}}{\cal A}^{\hat{b}}{\cal A}^{\hat{c}},
\end{align}
with\footnote{Physical $\widebar{27}^3$ Yukawa couplings have an overall factor $e^{K_{\rm cs}}$, 
whose factor is important to obtain the gauge invariant quantity under the symplectic group.}
\begin{align}
Y_{\hat{a}\hat{b}\hat{c}} &\equiv  e^{K_{\rm ks}}   (\Lambda^{-1/2}L)_{\hat{a}}^{d}  (\Lambda^{-1/2}L)_{\hat{b}}^{e} (\Lambda^{-1/2}L)_{\hat{c}}^{f}  \kappa_{def}.
\label{eq:physyukawa}
\end{align}

Note that not all the $27$ and $\widebar{27}$ representations behave as the chiral zero-modes in the 
low-energy effective action. 
The index theorem tells us that the net number of chiral zero-modes is  determined by the index
\begin{align}
\frac{\chi({\cal M})}{2}=h^{1,1}-h^{2,1},
\end{align}
corresponding to half the Euler characteristics of CY threefolds $\chi({\cal M})$. 
Hence, the chiral zero-modes in our interest are the remaining zero-modes originating from 
$27$ or $\widebar{27}$. 
Remarkably, the holomorphic Yukawa couplings in both $27$ and $\widebar{27}$ sectors are 
integers determined by the triple intersection numbers, 
indicating that it would be difficult to realize the hierarchical structure of holomorphic Yukawa couplings. 
Indeed, the difficulties of hierarchical structure of holomorphic Yukawa couplings 
on CY threefolds and toroidal orbifolds are discussed in detail in the following sections.

%%%%%%%%%%%%%%%%%%%%%%%%%%%%%%%%%%%%%%%%%%%%%%%%%%%%%%%%%%%%
\subsection{Role of K\"ahler metric on physical Yukawa couplings}
\label{subsec:2_2}
%%%%%%%%%%%%%%%%%%%%%%%%%%%%%%%%%%%%%%%%%%%%%%%%%%%%%%%%%%%%

In this section, we will focus on kinetic mixings to exemplify 
the difference between the physical Yukawa couplings and the 
holomorphic one on the basis of a simplified CY threefold in 
Section \ref{subsubsec:CY} and toroidal orbifolds in Section 
\ref{subsubsec:orbifolds}.

%%%%%%%%%%%%%%%%%%%%%%%%%%%%%%%%%%%%%%%%%%%%%%%%%%%%%%%%%%%%
\subsubsection{Calabi-Yau threefolds}
\label{subsubsec:CY}
%%%%%%%%%%%%%%%%%%%%%%%%%%%%%%%%%%%%%%%%%%%%%%%%%%%%%%%%%%%%

We start with CY threefolds to show the impacts of K\"ahler mixing on physical 
Yukawa couplings. 
It was known that typical CY threefolds contain hodge numbers with $|h^{1,1}-h^{2,1}|={\cal O}(100)$ 
as shown in Kreuzer-Skarke dataset of CY threefolds in toric ambient spaces \cite{Kreuzer:2000xy,Kreuzerdataset}, 
leading to ${\cal O}(100)$ number of chiral zero-modes. 
(See for several attempts to construct CY threefolds with a small hodge number, e.g., Ref.\cite{Candelas:2008wb}.) 

For the illustrative purpose, we assume that CY threefolds have the hodge number $h^{1,1}-h^{2,1}=3$ 
and the moduli space of remaining three-generation zero-modes is described by the 
special K\"ahler manifold $\left(\frac{SU(1,1)}{U(1)}\right)^3$. 
The K\"ahler potential of this geometry is described by
    \begin{align}
        K = -\sum_{a=1}^3 \ln (T^a + \widebar{T}^a),
    \end{align}
where $T^a$ denotes the K\"ahler moduli protected from becoming massive. 
The triple intersection number of these moduli is found as
\begin{align}
    \kappa_{123} = 1,
\end{align}
and 0 otherwise, and these correspond to the holomorphic $27^3$ Yukawa couplings. 
For concreteness, we identify the Higgs field with an element of $A^1$ and suppose that 
three generations of quarks and leptons appear from $A^{a}$ ($a=1,2,3$), associated with the 
massless K\"ahler moduli. 
Then, the holomorphic Yukawa couplings of quarks and leptons on the $(A^1,A^2,A^3)$ basis are of the form:
\begin{align}
\kappa_{1ab}=
\begin{pmatrix}
    0 & 0 & 0 \\
    0 & 0 & 1\\
    0 & 1 & 0 \\
\end{pmatrix}
\label{eq:holeg}
.
\end{align}
This rank-two matrix does not have non-trivial flavor and hierarchical structures. 
On these backgrounds, the K\"ahler metric is of the diagonal form due to the factorizable K\"ahler metric:
\begin{align}
K_{a\bar{b}} = \frac{\delta_{a\bar{b}}}{(T^a+\widebar{T}^a)^2},
\end{align}
yielding the non-vanishing element of physical Yukawa matrix
\begin{align}
Y_{\hat{1}\hat{2}\hat{3}}= e^{K_{\rm ks}} (\Lambda^{-1/2}L)_{\hat{1}}^{1}  (\Lambda^{-1/2}L)_{\hat{2}}^{2} (\Lambda^{-1/2}L)_{\hat{3}}^{3}\kappa_{123},
\end{align}
with the eigenvalue matrix $\Lambda$ and the diagonalizing matrix $L$ of moduli K\"ahler metric. 
Since the physical Yukawa couplings have the same structure as the holomorphic Yukawa couplings, 
it is still difficult to realize the non-trivial flavor and hierarchical structure of Yukawa couplings 
on such a special K\"ahler manifold with $\left(\frac{SU(1,1)}{U(1)}\right)^3$. 

When the K\"ahler moduli space differs from the $\left(\frac{SU(1,1)}{U(1)}\right)^3$ 
structure, a generic K\"ahler potential for the remaining 3 massless modes would be given by
    \begin{align}
        K = -\ln\bigl[ \kappa_{abc}(T^a + \widebar{T}^a) (T^b + \widebar{T}^b)(T^c + \widebar{T}^c) +\cdots \bigl],
    \end{align}
where the ellipsis represents contributions from other massive moduli. 
The matter kinetic mixings are thus induced by the intersection number $\kappa_{abc}$ 
and/or vacuum expectation values of heavy moduli, leading to the non-trivial flavor structure of $27$ zero-modes. 
The structure of physical Yukawa couplings (\ref{eq:physyukawa}) would be different from the holomorphic one due to 
the matter K\"ahler mixings. 
So far, we have focused on the K\"ahler moduli sector accompanying the $27$ 
zero-modes, but these statements hold for the opposite case with $h^{2,1}>h^{1,1}$, 
where $\widebar{27}$ zero-modes are in one-to-one correspondence with the complex structure moduli. 
Note that the holomorphic $\widebar{27}^3$ Yukawa couplings are also determined by triple intersection 
numbers $\kappa_{ijk}$ in the moduli space with the large complex structure. 

However, there is no definite way to analyze the CY moduli space with a large Euler number, 
although there exist the early three-generation models on a specific CY threefold taking into account freely acting discrete symmetries \cite{Greene:1986bm,Greene:1986jb}.
In this respect, we move on to toroidal orbifolds with and without resolutions, on which 
physical Yukawa couplings of untwisted and twisted modes are analyzed in detail below. 
The blown-up toroidal orbifolds would capture the structure of 
physical Yukawa couplings on smooth CY threefolds.

%%%%%%%%%%%%%%%%%%%%%%%%%%%%%%%%%%%%%%%%%%%%%%%%%%%%%%%%%%%%
\subsubsection{Toroidal orbifolds}
\label{subsubsec:orbifolds}
%%%%%%%%%%%%%%%%%%%%%%%%%%%%%%%%%%%%%%%%%%%%%%%%%%%%%%%%%%%%

The supersymmetric orbifold backgrounds are classified in Refs. \cite{Dixon:1986jc,Ibanez:1987pj,Font:1988mk}, which 
are summarized in Table \ref{tab:ZN} for $T^6/\mathbb{Z}_N$ and $T^6/(\mathbb{Z}_N\times \mathbb{Z}_M)$ orbifolds.
(See for Lie lattices of $T^6$ Refs.~\cite{Katsuki:1989bf,Kobayashi:1991rp} and for further classifications of orbifolds including non-Abelian ones, e.g., Ref. \cite{Fischer:2013qza}.)

\begin{table}%[H]
    \centering
    \begin{tabular}{|c|c|c|c|c|c|}\hline
       $\mathbb{Z}_N$  & Lattice & $h^{1,1}_{\rm untw.}$ & $h^{1,1}_{\rm twist}$ 
       & $h^{2,1}_{\rm untw.}$ & $h^{2,1}_{\rm twist}$ \\ \hline
        $\mathbb{Z}_3$ & $SU(3)^3$ & 9 & 27 & 0 & 0\\ \hline
        $\mathbb{Z}_4$ & $SU(2)^2 \times SO(5)^2$ & 5 & 26 & 1 & 6\\ 
         & $SU(2)\times SU(4) \times SO(5)$ & 5 & 22 & 1 & 2 \\ 
         & $SU(4)^2$ & 5 & 20 & 1 & 0\\ \hline
        $\mathbb{Z}_{6-I}$ & $SU(3)\times G_2^2$ & 5 & 24 & 0 & 5\\ 
         & $SU(3)^2\times G_2$ & 5 & 20 & 0 & 1\\ \hline
        $\mathbb{Z}_{6-II}$ & $SU(2)^2\times SU(3)\times G_2$ & 3 & 32 & 1 & 10\\ 
         & $SU(3)\times SO(8)$ & 3 & 26 & 1 & 4\\ 
         & $SU(2)^2\times SU(3)\times SU(3)$ & 3 & 28 & 1 & 6\\ 
         & $SU(2)\times SU(6)$ & 3 & 22 & 1 & 0\\ \hline
        $\mathbb{Z}_{7}$ & $SU(7)$ & 3 & 21 & 0 & 0\\ \hline
        $\mathbb{Z}_{8-I}$ & $SU(4)^2$ & 3 & 21 & 0 & 0\\ 
         & $SO(5)\times SO(9)$ & 3 & 24 & 0 & 3\\ \hline
        $\mathbb{Z}_{8-II}$ & $SU(2)\times SO(10)$ & 3 & 24 & 1 & 2\\ 
         & $SO(4)\times SO(9)$ & 3 & 28 & 1 & 6\\ \hline
        $\mathbb{Z}_{12-I}$ & $SU(3)\times F_4$ & 3 & 26 & 0 & 5\\ 
         & $E_6$ & 3 & 22 & 0 & 1\\ \hline
        $\mathbb{Z}_{12-II}$ & $SO(4)\times F_4$ & 3 & 28 & 1 & 6\\ \hline\hline
        $\mathbb{Z}_2\times \mathbb{Z}_2$ & $SU(2)^6$ & 3 & 48 & 3 & 0\\ \hline
        $\mathbb{Z}_2\times \mathbb{Z}_4$ & $SU(2)^2\times SO(5)^2$ & 3 & 58 & 1 & 0\\ \hline
        $\mathbb{Z}_2\times \mathbb{Z}_6$ & $SU(2)^2\times SU(3) \times G_2$ & 3 & 48 & 1 & 2\\ \hline
        $\mathbb{Z}_2\times \mathbb{Z}_{6^\prime}$ & $SU(3)\times G_2^2$  & 3 & 33 & 0 & 0\\ \hline
        $\mathbb{Z}_3\times \mathbb{Z}_3$ & $SU(3)^3$ & 3 & 81 & 0 & 0\\ \hline
        $\mathbb{Z}_3\times \mathbb{Z}_6$ & $SU(3)\times G_2^2$ & 3 & 70 & 0 & 1\\ \hline
        $\mathbb{Z}_4\times \mathbb{Z}_4$ & $SO(5)^3$ & 3 & 87 & 0 & 0\\ \hline
        $\mathbb{Z}_6\times \mathbb{Z}_6$ & $G_2^3$ & 3 & 81 & 0 & 0\\ \hline
    \end{tabular}
    \caption{Hodge number on toroidal orbifolds \cite{Ibanez:1987pj,Font:1988mk}.}
    \label{tab:ZN}
\end{table}

The K\"ahler potential of the untwisted moduli fields are categorized as follows \cite{Ferrara:1986qn,Cvetic:1988yw,Ibanez:1992hc}:
\begin{enumerate}
    \item $T^6/\mathbb{Z}_3$ orbifold
    \begin{align}
        K = -\ln \det (T^{a\bar{b}} + \widebar{T}^{a\bar{b}}), \quad 
        (a,b=1,2,3).
        \label{eq:Kahler1}
    \end{align}
    \item $T^6/\mathbb{Z}_{4,6}$ orbifolds
    \begin{align}
        K = -\ln (T^1 + \widebar{T}^1) -\ln \det (T^{a\bar{b}} + \widebar{T}^{a\bar{b}})
        -\sum_{n=1}^{h^{2,1}_{\rm untw.}} \ln (U^n + \widebar{U}^n), \quad 
        (a,b=2,3),
        \label{eq:Kahler2}
    \end{align}
    where the number of untwisted complex structure moduli is listed in Table \ref{tab:ZN}.
    \item $T^6/\mathbb{Z}_{7,8,12}$ and $T^6/(\mathbb{Z}_N\times \mathbb{Z}_M)$ orbifolds
    \begin{align}
        K = -\sum_{a=1}^3 \ln (T^a + \widebar{T}^a) 
        -\sum_{n=1}^{h^{2,1}_{\rm untw.}} \ln (U^n + \widebar{U}^n),
        \label{eq:Kahler3}
    \end{align}
    where the number of untwisted complex structure moduli is listed in Table \ref{tab:ZN}.
\end{enumerate}
The number of the K\"ahler moduli is larger than that of the complex structure moduli 
in all the toroidal orbifolds, and consequently the chiral zero-modes are originating from the K\"ahler moduli.\footnote{It is also 
possible to consider the opposite case, that is, $h^{2,1}>h^{1,1}$ by choosing the proper assignment of 
discrete torsion \cite{Font:1988mk}. 
Indeed, our analysis is applicable to that case as demonstrated on a mirror dual of the 
$T^6/(\mathbb{Z}_3\times \mathbb{Z}_3)$ orbifold in Section \ref{subsec:orbifold3}.} 
In this respect, we examine the holomorphic $27^3$ Yukawa couplings.

Let us study the holomorphic $27^3$ Yukawa coupling associated with the untwisted K\"ahler moduli 
on $T^6/\mathbb{Z}_{7,8,12}$ and $T^6/(\mathbb{Z}_N\times \mathbb{Z}_M)$ orbifolds, on which the non-vanishing triple intersection number is evaluated as
\begin{align}
    \kappa_{123} = 1.
\end{align}
Hence, the untwisted K\"ahler moduli space on these toroidal orbifolds is described by the 
special K\"ahler manifold with $\left(\frac{SU(1,1)}{U(1)}\right)^3$. 
When we  identify the Higgs field and three generations of quarks/leptons with, for instance, 
the elements of $A^1$ and $A^{1,2,3}$ associated with the untwisted K\"ahler moduli, 
the holomorphic Yukawa couplings of quarks and leptons become the rank-two matrix. 
There is no significant difference between the physical Yukawa couplings and 
the holomorphic one as analyzed before. 
In this way, it is difficult to realize the hierarchical structure of Yukawa couplings from the untwisted K\"ahler moduli. Similar phenomena occur in the $\widebar{27}^3$ Yukawa couplings associated with the untwisted complex structure moduli on  all the toroidal orbifolds whose K\"ahler potential is enumerated in Eqs. (\ref{eq:Kahler1})-(\ref{eq:Kahler3}).

One of the possibilities to realize the difference between the holomorphic Yukawa couplings and the physical one is 
to incorporate the off-diagonal K\"ahler metric of matter fields. 
The matter kinetic mixings are sourced by (i) the unfactorizable toroidal orbifolds 
such as $\mathbb{Z}_{3,4,6}$ orbifolds and/or (ii) the twisted modes in the effective action. 
In the next section, we study both possibilities to clarify the crucial role of off-diagonal 
entries in matter field K\"ahler metric. 
Before concluding this section, we exhibit a constraint on the moduli  K\"ahler metric in the analysis of 
$27^3$ Yukawa couplings.

In the context of heterotic string theory, the value of 4D gauge coupling is directly 
related to the size of internal volume. 
We recall that the ten-dimensional (10D) heterotic supergravity action in string frame for our interest is of the form:
\begin{align}
    S = \frac{M_{10}^8}{2}\int d^{10}x \sqrt{-G}e^{-2\phi}R - \frac{1}{2g_{10}^2}\int d^{10}x \sqrt{-G}e^{-2\phi}{\rm Tr}F^2,
\end{align}
where $F$ denotes the gauge field strength of $E_8\times E_8$ or $SO(32)$ gauge group with the adjoint representation. 
The 10D gravitational and gauge couplings are related as
\begin{align}
    M_{10}^8= \frac{4\pi}{l_s^8}=\frac{4}{\alpha^\prime g_{10}^2},
\end{align}
with the string length $l_s=2\pi\sqrt{\alpha^\prime}$. 
When we compactify the theory on a 6D internal manifold ${\cal M}$ with the volume ${\rm Vol}({\cal M})={\cal V}l_s^6$ measured in units of the string length, the 4D gauge coupling is determined by the internal volume ${\cal V}$ and the string coupling 
$g_s=e^{\langle \phi \rangle}$,
\begin{align}
g_4^{-2} = e^{-2\langle \phi \rangle}{\rm Vol}({\cal M})g_{10}^{-2} = \frac{{\cal V}}{4\pi g_s^{2}}.
\end{align}
It indicates that the internal volume should obey
\begin{align}
    {\cal V}= g_s^2\alpha^{-1} \lesssim 25,
\end{align}
where we impose the small string coupling $g_s\leq 1$ and the 4D gauge coupling $\alpha^{-1}=4\pi g_4^{-2}\simeq 25$ required in supersymmetric $SU(5)$ grand unified theory. 
Supersymmetry breaking at a high energy scale may allow larger $\alpha^{-1}$.
In the following analysis, 
we allow for ${\cal V}\leq 30$ to be applicable to 
a broad class of 4D low-energy models.\footnote{See Ref. \cite{Abe:2015xua} for one-loop contributions to the gauge couplings 
in phenomenologically promising models from heterotic string theory.}
A strong correlation between the 4D gauge coupling and the size of internal volume is well known in a 
global model such as the heterotic string theory, 
but we will argue that the limited range of internal volume 
also constrain the structure of physical $27^3$ Yukawa couplings as discussed in detail later.

%%%%%%%%%%%%%%%%%%%%%%%%%%%%%%%%%%%%%%%%%%%%%%%%%%%%%%%%%%%%
%%%%%%%%%%%%%%%%%%%%%%%%%%%%%%%%%%%%%%%%%%%%%%%%%%%%%%%%%%%%
\section{Hierarchical structure of physical Yukawa couplings}
\label{sec:3}
%%%%%%%%%%%%%%%%%%%%%%%%%%%%%%%%%%%%%%%%%%%%%%%%%%%%%%%%%%%%
%%%%%%%%%%%%%%%%%%%%%%%%%%%%%%%%%%%%%%%%%%%%%%%%%%%%%%%%%%%%

In this section, we explicitly demonstrate the hierarchical structure of physical 
Yukawa couplings on three backgrounds such as 
(i) $T^6/\mathbb{Z}_3$ orbifold without twisted modes in Section \ref{subsec:orbifold1}, 
(ii) $T^6/(\mathbb{Z}_2\times \mathbb{Z}_2)$ orbifold with twisted modes in Section \ref{subsec:orbifold2}, 
(iii) a mirror dual of $T^6/(\mathbb{Z}_3\times \mathbb{Z}_3)$ orbifold with twisted modes in Section \ref{subsec:orbifold3},
respectively.

%%%%%%%%%%%%%%%%%%%%%%%%%%%%%%%%%%%%%%%%%%%%%%%%%%%%%%%%%%%%
\subsection{$T^6/\mathbb{Z}_3$ orbifold}
\label{subsec:orbifold1}
%%%%%%%%%%%%%%%%%%%%%%%%%%%%%%%%%%%%%%%%%%%%%%%%%%%%%%%%%%%%

For the first illustrative example, we deal with the $T^6/\mathbb{Z}_3$ orbifold, 
where the moduli fields on $T^6/\mathbb{Z}_3$ geometry consist of 9 untwisted and 27 twisted K\"ahler moduli.  
It is known that the K\"ahler potential of the untwisted K\"ahler moduli described by 
the  $\frac{SU(3,3)}{SU(3)\times SU(3)\times U(1)}$ coset space is provided by
\begin{align}
    K_{\rm ks}=  -\ln \det (T + \widebar{T}),
\end{align}
with 
\begin{align}
T=
\begin{pmatrix}
T^1 & T^4 & T^5\\
T^7 & T^2 & T^6\\
T^8 & T^9 & T^3\\
\end{pmatrix}
,
\end{align}
and the corresponding triple intersection numbers are given by
\begin{align}
    \kappa_{123}=\kappa_{468}=\kappa_{579}=1, \quad \kappa_{169}=\kappa_{258}=\kappa_{347}=-1,
\end{align}
and 0 otherwise. 
The matter K\"aher potential and corresponding metric are also known as\cite{Cvetic:1989ii}
\begin{align}
    K &= -\ln \det (T + \widebar{T}-A^\alpha A^{\bar{\alpha}})^{a\bar{b}} 
    = -\ln \det (T + \widebar{T})^{a\bar{b}} + (T + \widebar{T})^{-1}_{a\bar{b}}A^a_\alpha A^{\bar{b}}_{\bar{\alpha}} +{\cal O}(|A|^4),
    \nonumber\\
    K_{a\bar{b}}^{(27)} &= (T + \widebar{T})^{-1}_{a\bar{b}},
\end{align}
where $\alpha$ denotes the indices of $SU(3)\subset E_8$. Note that the indices of $\alpha$ and $a$ are not identified with each other. 
The superpotential takes the form\footnote{Since we are interested in the realization of hierarchical structure of physical Yukawa couplings, we simply omit the overall factor of the superpotential in the subsequent calculations.}
\begin{align}
W = \kappa_{abc}\epsilon^{\alpha \beta \gamma}A^a_\alpha A^b_\beta A^c_\gamma,
\end{align}
where $ \kappa_{abc}$ and $\epsilon^{\alpha \beta \gamma}$ correspond to the anti-symmetric tensors for $SU(3)$ isometry  and $SU(3)\subset E_8$, respectively.

In contrary to the previous analysis, there exist the non-vanishing matter kinetic mixing. 
To examine the effects of off-diagonal entries in matter K\"ahler metric to the physical Yukawa couplings, 
we focus on $27^3$ Yukawa couplings of untwisted modes $\{A^1_1,A^2_2,A^3_3\}$ by assuming that they are protected 
from becoming massive. Note that the following analysis is applicable to other untwisted modes and we rewrite $\{A^1_1,A^2_2,A^3_3\}$ as $\{A^1,A^2,A^3\}$ to shorten the notation. 
Furthermore, we restrict ourselves on the locus $T^4=T^5=T^6=T^7=T^8=T^9$ to simplify our analysis. 
On this locus, the moduli K\"ahler potential and matter K\"ahler metric are evaluated as follows:
\begin{align}
   K_{\rm ks}=& -\ln \Bigl[(T^1+\widebar{T}^1)(T^2+\widebar{T}^2)(T^3+\widebar{T}^3) -
    \sum_{a=1}^3 (T^a+\widebar{T}^a)(T^4+\widebar{T}^4)^2 +2(T^4+\widebar{T}^4)^3 \Bigl],
    \nonumber\\
K_{a\bar{b}}^{(27)}=&\frac{1}{2t^1t^2t^3 - 2(t^1 + t^2 + t^3) (t^4)^2 + 4 (t^4)^3}
\begin{pmatrix}
t^2t^3 - (t^4)^2 & 
(t^4 - t^3) t^4 & 
(t^4 - t^2) t^4 \\
(t^4 - t^3) t^4 & 
t^1t^3 - (t^4)^2 & 
(t^4 - t^1) t^4 \\
(t^4 - t^2) t^4 &
(t^4 - t^1) t^4 &
t^1t^2 - (t^4)^2 \\
\end{pmatrix}
,
\label{eq:KahlerZ3}
\end{align}
with $t^a={\rm Re}(T^a)$, 
which reduces to the diagonal matter K\"ahler metric in the $t^4 \rightarrow 0$ limit, 
namely the value of $t^4$ controls whether the K\"ahler metric is diagonal.

When the vacuum expectation values of the untwisted K\"ahler moduli are restricted along 
the isotropic locus,
\begin{align}
\langle t^1\rangle= \langle t^2\rangle =\langle  t^3\rangle =x,\quad \langle t^4\rangle=x^p,
\label{eq:vev1}
\end{align}
with $x>1$ and $p\geq 0$,\footnote{We restrict ourselves to these moduli space, otherwise the value of $t^{1,2,3,4}$ can be smaller than 1 in contradiction with the reliability of the supergravity description.} eigenvalues of the moduli K\"ahler metric are simply obtained as
\begin{align}
\Lambda_{\hat{a}\bar{\hat{b}}} =&
\begin{pmatrix}
(2x -2x^p)^{-1} & 
0 & 
0 \\
0 & 
(2x -2x^p)^{-1} & 
0 \\
0 &
0 &
(2x +4x^p)^{-1}\\
\end{pmatrix}
\label{eq:LambdaZ3iso}
\end{align}
by employing the following diagonalizing matrix of moduli K\"ahler metric: 
\begin{align}
L=&
\begin{pmatrix}
-1 & 
0 & 
1 \\
-1 & 
1 & 
0 \\
1 &
1 &
1 \\
\end{pmatrix}
.
\end{align}
The physical Yukawa couplings are then computed by substituting these into the 
formula (\ref{eq:physyukawa}). 
It turns out that the physical Yukawa couplings $Y_{\hat{1}\hat{a}\hat{b}}$ and $Y_{\hat{2}\hat{a}\hat{b}}$ become 
the rank-two matrices due to the presence of degenerate eigenvalues in the K\"ahler metric (\ref{eq:LambdaZ3iso}). 
For that reason, we assume that the Higgs field is originating from $A^{\hat{3}}$, 
and quarks/leptons are the elements of $(A^{\hat{1}},A^{\hat{2}},A^{\hat{3}})$, respectively. 
Indeed, from an analytical expression of physical Yukawa couplings $Y_{\hat{3}\hat{a}\hat{b}}$, 
\begin{align}
Y_{\hat{3}\hat{a}\hat{b}} &= e^{K_{\rm ks}} \sum_{d,e,f=1}^3 (\Lambda^{-1/2}L)_{\hat{3}}^{d}  (\Lambda^{-1/2}L)_{\hat{b}}^{e} (\Lambda^{-1/2}L)_{\hat{c}}^{f} \kappa_{def}
\nonumber\\
&=e^{K_{\rm ks}} (2\sqrt{2})^{-1}(x+2x^p)^{-1/2}
\begin{pmatrix}
 2(-x+x^p) & 
   -x+x^p & 
0 \\
-x+x^p & 
2(-x+x^p) & 
0\\
0 &
0 &
6(x+2x^p) \\
\end{pmatrix}
,
\end{align}
the determinant
\begin{align}
\det(Y_{\hat{3}\hat{a}\hat{b}}) &= e^{3K_{\rm ks}}288\sqrt{2}(x-x^p)^2 (x+2x^p)^{5/2}
\end{align}
turns out to be non-zero unless $p=0$. 
One can realize the rank-three physical Yukawa matrix unless $p= 0$, 
although the holomorphic Yukawa coupling (\ref{eq:holeg}) is rank-two matrix. 
Hence, off-diagonal entries in the matter field K\"ahler metric bring us a rich 
structure of the physical Yukawa couplings. 
In the current parametrization of the moduli fields (\ref{eq:vev1}), 
eigenvalues of $Y_{\hat{3}\hat{a}\hat{b}}$ are calculated as
\begin{align}
\begin{split}
m_1 &= e^{K_{\rm ks}}12\sqrt{2}(x+2x^p)^{3/2},
\nonumber\\
m_2 &= e^{K_{\rm ks}}6\sqrt{2}(-x+x^p)(x+2x^p)^{1/2},
\nonumber\\
m_3 &= e^{K_{\rm ks}}2\sqrt{2}(-x+x^p)(x+2x^p)^{1/2},
\end{split}
\end{align}
with the order $|m_1|>|m_2|>|m_3|$ and these ratios become
\begin{align}
r_1 = \frac{|m_2|}{|m_1|}=\frac{|x-x^p|}{2(x+2x^p)},\quad
r_2=\frac{|m_3|}{|m_1|}=\frac{|x-x^p|}{6(x+2x^p)},
\end{align}
satisfying $|m_3|/|m_2|=1/3$. 
Taking into account the constraint for the internal volume 
\begin{align}
    {\cal V}= 8(x - x^p)^2 (x + 2 x^p)\leq 30,
\end{align} 
it turns out that the typical ratio $|m_3|/|m_1|$ is of ${\cal O}(0.01)$ to be consistent with the realistic value of 4D gauge coupling.

To exemplify the realization of  the hierarchical physical Yukawa couplings, 
we search the anisotropic untwisted K\"ahler moduli by changing the moduli locus (\ref{eq:vev1}) 
to 
\begin{align}
\langle t^1\rangle= \langle t^2\rangle =x, \quad 
\langle  t^3\rangle =y,\quad \langle t^4\rangle=z,
\end{align}
with $x,y,z>1$. 
When we adopt the above parametrization of the moduli fields, the moduli K\"ahler metric (\ref{eq:KahlerZ3}) is diagonalized as follows:
\begin{align}
\Lambda_{\hat{a}\bar{\hat{b}}} =&
\begin{pmatrix}
\frac{1}{2(x-z)} & 
0 & 
0 \\
0 & 
\Lambda_-^{-1} & 
0 \\
0 &
0 &
\Lambda_+^{-1}\\
\end{pmatrix}
,
\end{align}
with 
\begin{align}
\begin{split}
        \Lambda_\pm&= 
   x + y + z \pm \sqrt{x^2 + y^2 + 2 x z + 9 z^2 - 2 y (x + z)}.
\end{split}
\end{align}
Here, the diagonalization of the K\"ahler metric is performed by
\begin{align}
L=&
\begin{pmatrix}
-1 & 
1 & 
0 \\
L_+ & 
L_+ & 
1 \\
L_- & 
L_- & 
1 \\
\end{pmatrix}
,
\label{eq:LZ3aniso}
\end{align}
with
\begin{align}
    L_\pm = \frac{x - y + z \pm \sqrt{x^2 + y^2 + 2 x z + 9 z^2 - 2 y (x + z)}}{4 z}.
\end{align}
We arrive at the physical Yukawa couplings by substituting 
these quantities into Eq. (\ref{eq:physyukawa}). 
Since the analytical expression of the physical Yukawa couplings is complicated with respect to the moduli fields, we rely on the numerical search to find out desirable hierarchical structure of physical Yukawa couplings. 
In particular, we focus on physical Yukawa coupling $Y_{\hat{3}\hat{a}\hat{b}}$, 
setting one of moduli fields at specific values. 
Note that $Y_{\hat{1}\hat{a}\hat{b}}$ becomes the rank-two matrix 
and $Y_{\hat{2}\hat{a}\hat{b}}$ has the same structure as  $Y_{\hat{3}\hat{a}\hat{b}}$ due to the structure of 
diagonalizing matrix (\ref{eq:LZ3aniso}).

\begin{figure}[H]
 \begin{minipage}{0.49\hsize}
  \begin{center}
    \includegraphics[scale=0.5]{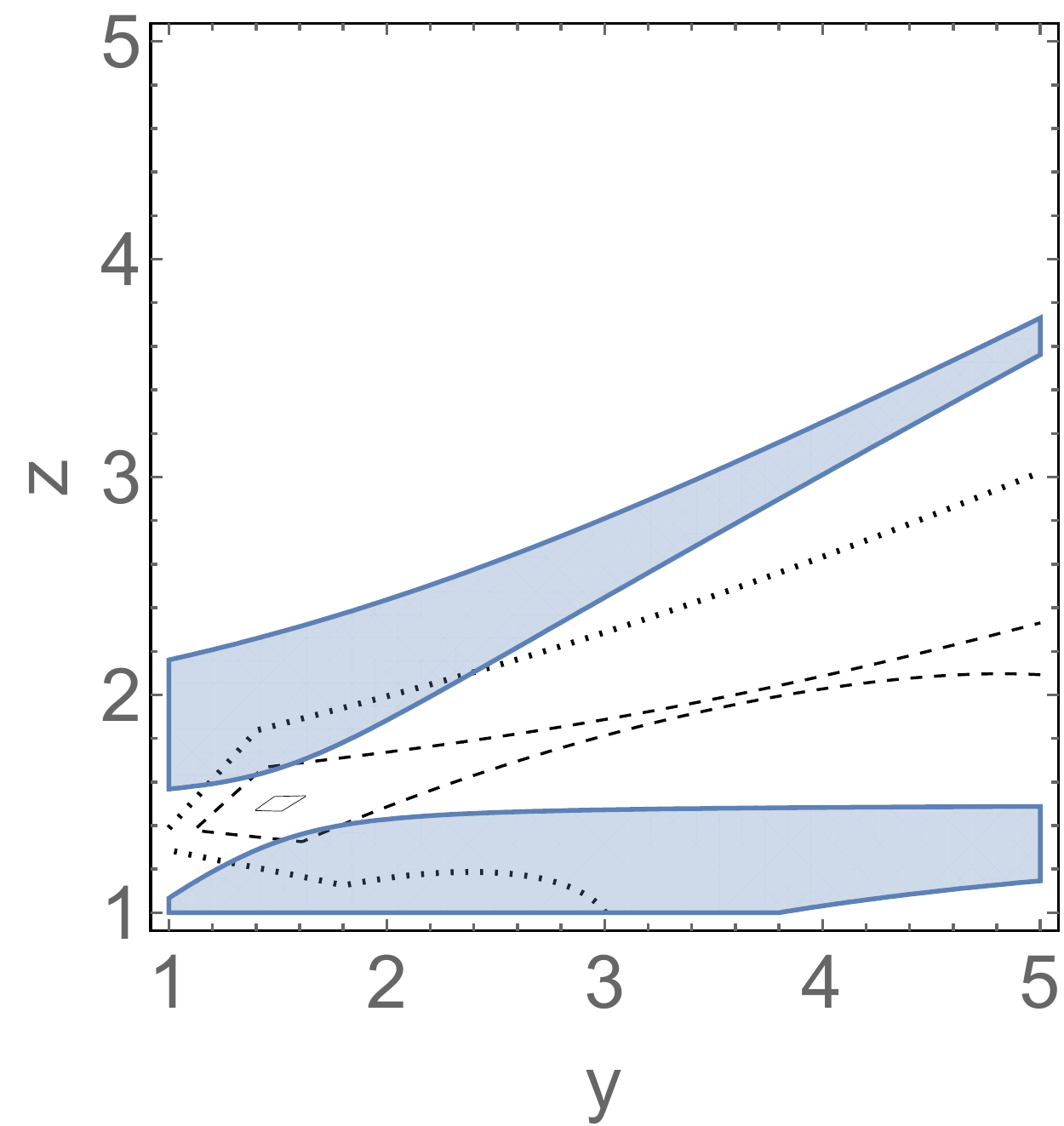}
  \end{center}
 \end{minipage}
 \begin{minipage}{0.49\hsize}
  \begin{center}
    \includegraphics[scale=0.5]{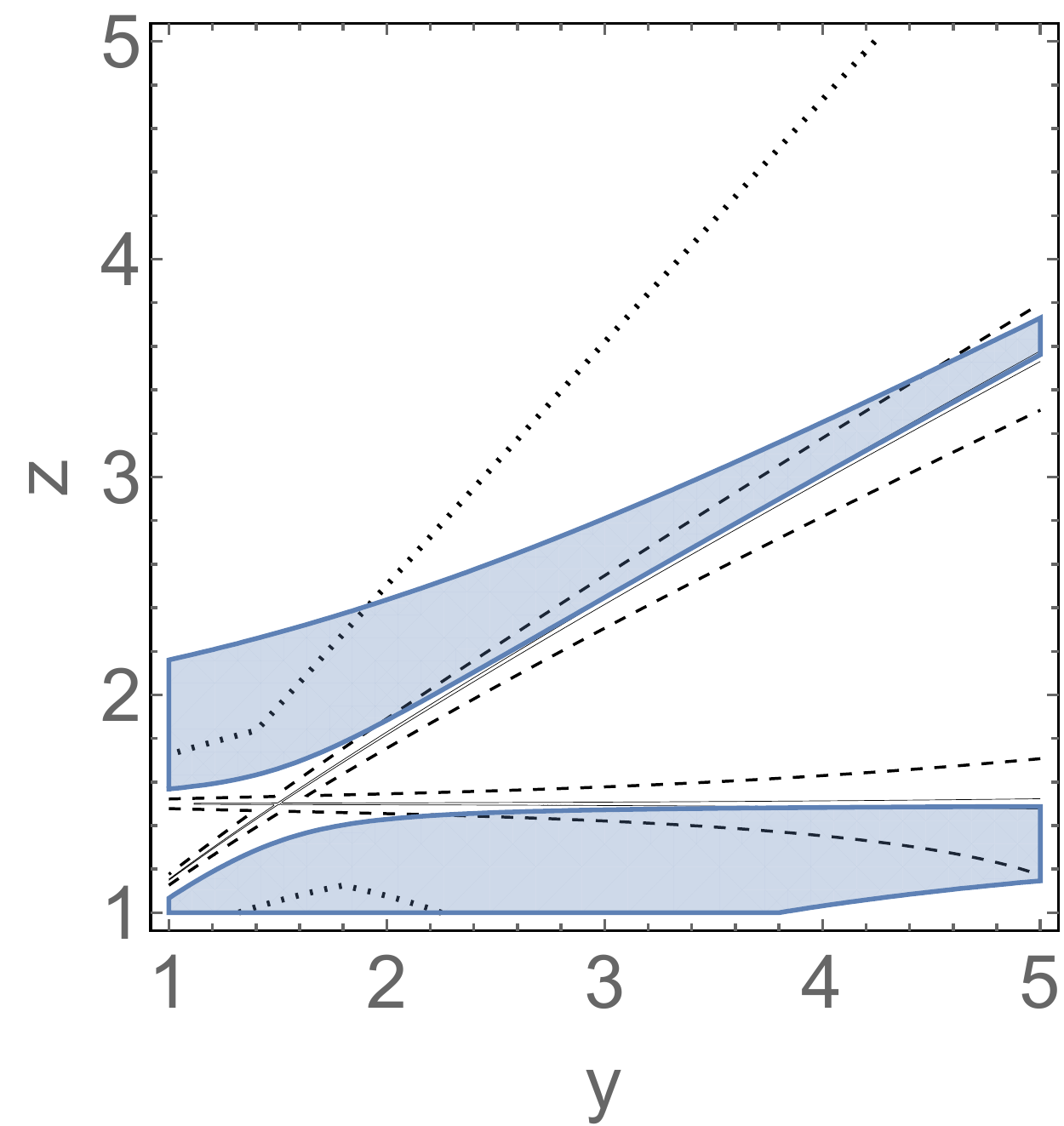}
  \end{center}
 \end{minipage}
\caption{In the left and right panels, we plot two ratios $r_1$ and $r_2$ of eigenvalues of the physical Yukawa couplings ($Y_{\hat{3}\hat{a}\hat{b}}$) to the maximum one at the fixed $x=1.5$, respectively.  
The dotted, dashed, and solid curves correspond to the values for $10^{-1}$ ($10^{-1}$), $0.05$  $(10^{-2})$, and $10^{-2}$ $(10^{-3})$, in the left (right) panel. 
In both panels, blue shaded regions represent the allowed volume $1<{\cal V}\leq 30$. 
Note that the Yukawa matrix reduces to the rank-two Yukawa matrix at $z=\frac{y + \sqrt{y(8x+y)}}{4}$ and $z=x$, in the vicinity of the solid curves.}
    \label{fig:aniiso1}
\end{figure}

\begin{figure}[H]
 \begin{minipage}{0.49\hsize}
  \begin{center}
    \includegraphics[scale=0.5]{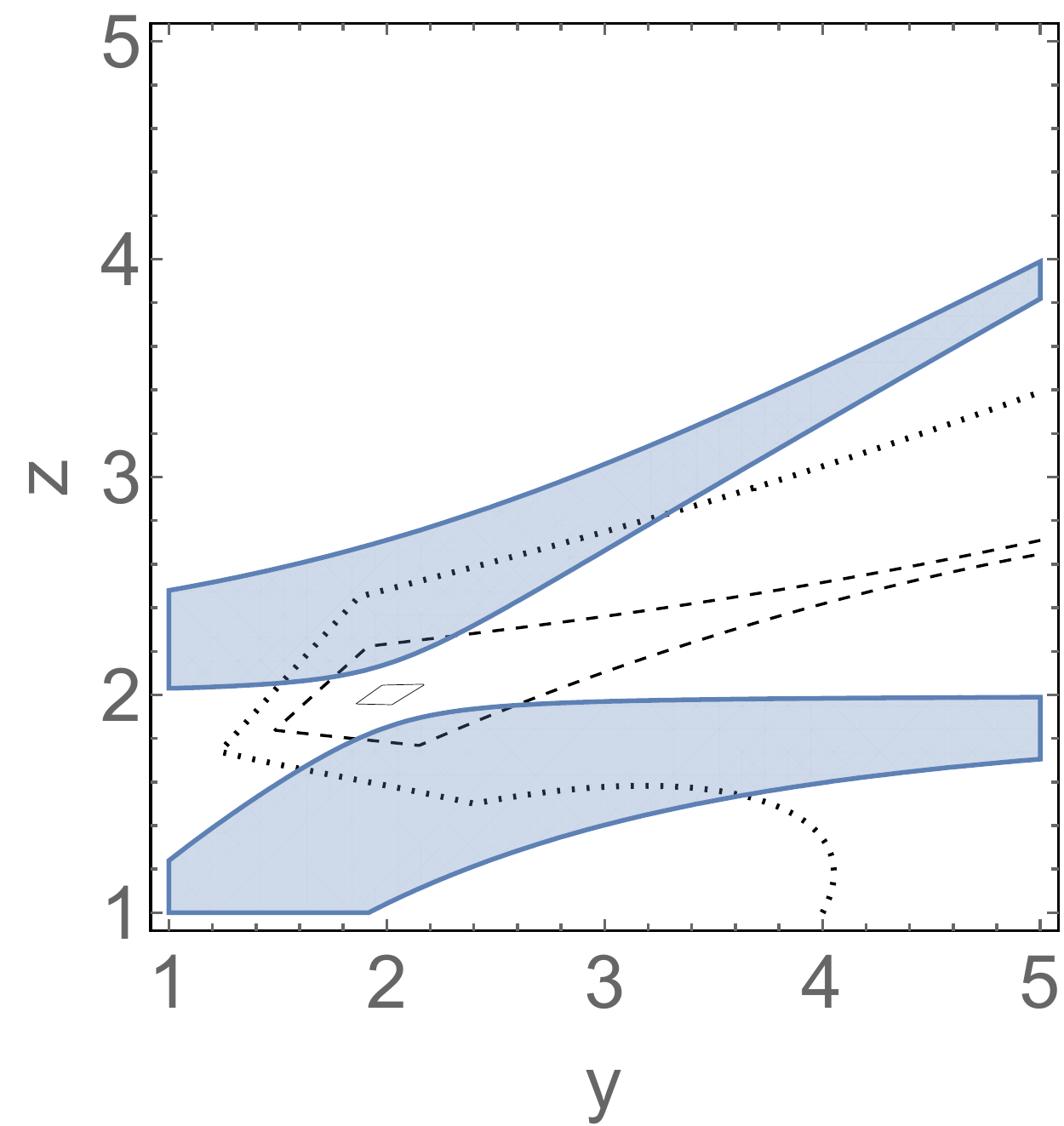}
  \end{center}
 \end{minipage}
 \begin{minipage}{0.49\hsize}
  \begin{center}
    \includegraphics[scale=0.5]{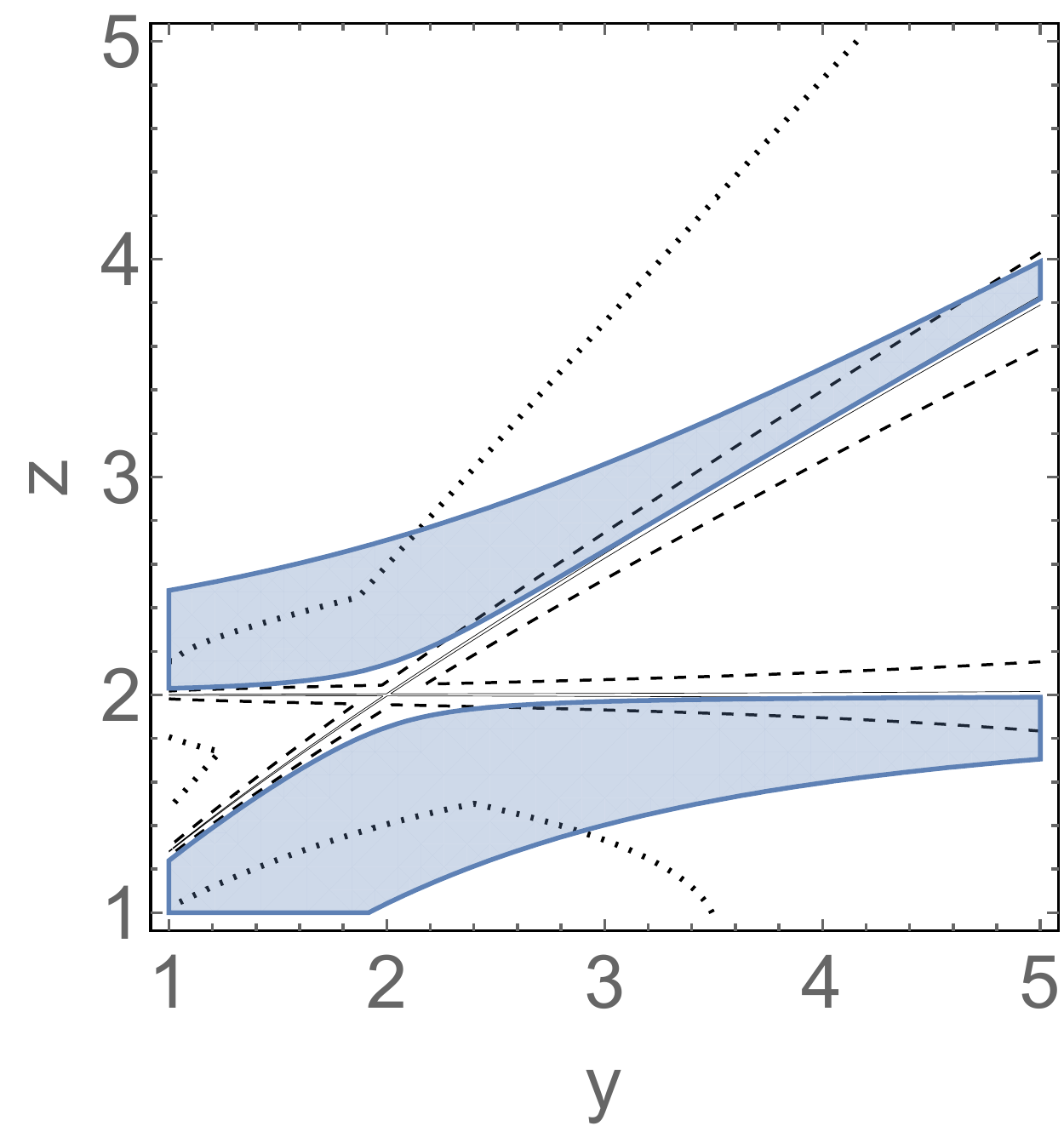}
  \end{center}
 \end{minipage}
\caption{In the left and right panels, we plot two ratios $r_1$ and $r_2$ of eigenvalues of the physical Yukawa couplings ($Y_{\hat{3}\hat{a}\hat{b}}$) to the maximum one at the fixed $x=2$, respectively.  
The dotted, dashed, and solid curves correspond to the values for $10^{-1}$ ($10^{-1}$), $0.05$  $(10^{-2})$, and $10^{-2}$ $(10^{-3})$, in the left (right) panel. 
In both panels, blue shaded regions represent the allowed volume $1<{\cal V}\leq 30$. Note that the Yukawa matrix reduces to the rank-two Yukawa 
matrix at $z=\frac{y + \sqrt{y(8x+y)}}{4}$ and $z=x$, in the vicinity of the solid curves.}
    \label{fig:aniiso2}
\end{figure}

We plot two ratios of eigenvalues of the physical Yukawa couplings ($Y_{\hat{3}\hat{a}\hat{b}}$) to the maximum one 
in the left panel for $r_1$ and the right panel for $r_2$ in Figure \ref{fig:aniiso1} with $x=1.5$ 
and Figure \ref{fig:aniiso2} with $x=2$. 
In contrast to the case with isotropic untwisted K\"ahler moduli, 
the solid curves in Figures \ref{fig:aniiso1} and \ref{fig:aniiso2} show that 
${\cal O}(1)$ values of the moduli fields  induce the ${\cal O}(10^{-2})$ and ${\cal O}(10^{-3})$ hierarchical 
structures of the physical Yukawa couplings for $r_1$ and $r_2$.
A tuning of the moduli fields leads to the more hierarchical structure of physical Yukawa couplings, 
irrespective of the value $x$.

Remarkably, there exists an overlap region between the region with the hierarchical Yukawa 
couplings and the desired value of the toroidal volume ${\cal V}\leq 30$. 
Note that the volume of the toroidal orbifold ${\cal V}=8(x-z)\left( xy+(y-2z)z\right)$ goes to zero when $y$ and $z$ approach to $x$. 
In Table \ref{tab:bench}, we show three benchmark values for the ratio $(1, r_1,r_2)$ at 
several values of the K\"ahler moduli.

\begin{table}[H]
    \centering
    \begin{tabular}{|c|c|c|}\hline
     $(x,\,y,\,z)$     &  $(1,\,r_1,\,r_2)$ & ${\cal V}$ \\ \hline
      (1.5, 4, 3.2)   & $(1, 0.15, 1.1\times 10^{-2})$ & 22.8 \\ \hline
      (1.5, 5, 3.7)   & $(1, 0.15, 6.0\times 10^{-3})$ & 24.3 \\ \hline      
      (2, 4.5, 3.7)   & $(1, 0.14, 9.2\times 10^{-3})$ & 23.5 \\ \hline     
    \end{tabular}
    \caption{Benchmark ratios for the ratio $(1,r_1,r_2)$ and the toroidal volume ${\cal V}$ at 
    several moduli values.}
    \label{tab:bench}
\end{table}

\subsection{Blown-up $T^6/(\mathbb{Z}_2\times \mathbb{Z}_2)$ orbifold}
\label{subsec:orbifold2}
%%%%%%%%%%%%%%%%%%%%%%%%%%%%%%%%%%%%%%%%%%%%%%%%%%%%%%%%%%%%

In this section, we discuss the $T^6/(\mathbb{Z}_2\times \mathbb{Z}_2)$ orbifold 
including twisted modes. 
On the $T^6/(\mathbb{Z}_2\times \mathbb{Z}_2)$ geometry, 
the K\"ahler metric of matter fields associated with the untwisted K\"ahler 
moduli is factorizable and their physical Yukawa couplings have the same 
structure of the holomorphic Yukawa couplings. 
Hence, we study in detail the impacts of twisted K\"ahler moduli on the 
physical Yukawa coupling. 

The K\"ahler form on the blown-up $T^6/(\mathbb{Z}_2\times \mathbb{Z}_2)$ geometry is expanded as
\begin{align}
    J = \sum_{a=1}^3 t^a R_a -\sum_{r=1}^{48}s^rE_r,
\end{align}
where $t^a$ denotes the untwisted K\"ahler moduli associated 
with divisors $R_a$ represented as $dz_i\wedge d\bar{z}_i$ using the complex coordinates of tori $z_i$, 
whereas $s^r$ denote the twisted K\"ahler moduli associated with exceptional divisors $E_r$. 
Note that $t^a$ and $s^r$ are chosen to be positive to give a geometrical interpretation, and 
furthermore $t^a$ should be larger than $s^r$ to ensure the positivity of curves, divisors and 
entire volume.

In this paper, we restrict ourselves to the simplified case, where all the twisted K\"ahler moduli 
are identified with
\begin{align}
    t^4\equiv s^r,\quad (r=1,2,\cdots,48).
\end{align}
Here and in what follows, we follow the result in Eq. (6.8) of Ref. \cite{Denef:2005mm}, although 
there exists a huge possibility of resolutions. (For more details, see, Refs. \cite{Denef:2005mm,Blaszczyk:2010db}.) 
The K\"ahler potential on resolutions of the blown-up $T^6/(\mathbb{Z}_2\times \mathbb{Z}_2)$ orbifold 
is given by
\begin{align}
    K=-\ln \Bigl[\frac{1}{8}(T^1+\widebar{T}^1)(T^2+\widebar{T}^2)(T^3+\widebar{T}^3) -
    \sum_{a=1}^3 (T^a+\widebar{T}^a)(T^4+\widebar{T}^4)^2 +6(T^4+\widebar{T}^4)^3\Bigl], 
\end{align}
with $t^a ={\rm Re}(T^a)$, 
approaching to the volume of $T^6/(\mathbb{Z}_2\times \mathbb{Z}_2)$ orbifold 
in the limit $t^{1,2,3}\gg t^4$. 
Here the effective triple intersection numbers are 
\begin{align}
    \kappa_{123}=1,\quad \kappa_{a44}=-16,\quad \kappa_{444}=288,
\end{align}
with $a=1,2,3$, and 0 otherwise. 
The inclusion of the blow-up modes gives rise to the matter kinetic mixing. 
Indeed, the K\"ahler potential has a similar structure to $T^6/\mathbb{Z}_3$ orbifold 
with the identification $T^4=T^5=T^6=T^7=T^8=T^9$ as shown in Eq. (\ref{eq:KahlerZ3}), 
where only the difference is the coefficient of $(T^4+\widebar{T}^4)$. 
In the following subsections, we assume the specific modes protected becoming massive 
to reveal the structure of physical Yukawa couplings. 
In particular, we analyze two types of Yukawa couplings: 
(i) three-point couplings of untwisted modes themselves in Section \ref{subsubsec:untw} and 
(ii) three-point couplings of (untwisted)-(untwisted)-(twisted) modes in Section \ref{subsubsec:tw}.

%%%%%%%%%%%%%%%%%%%%%%%%%%%%%%%%%%%%%%%%%%%%%%%%%%%%%%%%%%%%
\subsubsection{Yukawa couplings of untwisted modes}
\label{subsubsec:untw}
%%%%%%%%%%%%%%%%%%%%%%%%%%%%%%%%%%%%%%%%%%%%%%%%%%%%%%%%%%%%

To analyze the structure of physical Yukawa couplings of untwisted modes themselves, 
we search the anisotropic untwisted K\"ahler moduli:
\begin{align}
\langle t^1\rangle= \langle t^2\rangle =x, \quad 
\langle  t^3\rangle =y,\quad \langle t^4\rangle=z,
\label{eq:vev2}
\end{align}
with $x,y,z>1$. 
When we adopt the above parametrization of the moduli fields, the moduli K\"ahler metric is calculated as
\begin{align}
\Lambda_{\hat{a}\bar{\hat{b}}}^{(\mathbb{Z}_2\times \mathbb{Z}_2)} =&\frac{1}{8(x^2y-8(2x+y)z^2+48z^3)^2}
\begin{pmatrix}
2 y (x^2 y - 8 (2 x + y) z^2 + 48 z^3) & 
0 & 
0 \\
0 & 
\Lambda_-^{(\mathbb{Z}_2\times \mathbb{Z}_2)} & 
0 \\
0 &
0 &
\Lambda_+^{(\mathbb{Z}_2\times \mathbb{Z}_2)}\\
\end{pmatrix}
,
\end{align}
with 
\begin{align}
\begin{split}
        \Lambda_\pm^{(\mathbb{Z}_2\times \mathbb{Z}_2)}&= 
   x^4 - 16 x y z^2 + x^2 (y^2 - 16 z^2) + 8 z^2 (y^2 - 6 y z + 24 z^2)\pm \sqrt{\Lambda^{(\mathbb{Z}_2\times \mathbb{Z}_2)}},
    \nonumber\\
\Lambda^{(\mathbb{Z}_2\times \mathbb{Z}_2)} &\equiv -2048 x (3 x - 8 z) (x - 3 z) z^5 + 
 64 y z^2 (x + 3 z) (x^2 - 8 z^2)^2 - 
 4 y^2 (x^2 - 8 z^2)^2 (x^2 + 8 z^2)
 \nonumber\\
 &+ (x^4 - 16 x^2 z^2 + 192 z^4 - 
   16 y z^2 (x + 3 z) + y^2 (x^2 + 8 z^2))^2.
\end{split}
\end{align}
Here, the diagonalization of the K\"ahler metric is performed by
\begin{align}
L^{(\mathbb{Z}_2\times \mathbb{Z}_2)}=&
\begin{pmatrix}
-1 & 
1 & 
0 \\
L_1-\frac{\Lambda_+^{(\mathbb{Z}_2\times \mathbb{Z}_2)}}{32 z^2 (8 y^2 - 6 z x + x^2)} & 
L_1-\frac{\Lambda_+^{(\mathbb{Z}_2\times \mathbb{Z}_2)}}{32 z^2 (8 y^2 - 6 z x + x^2)} & 
1 \\
L_1-\frac{\Lambda_-^{(\mathbb{Z}_2\times \mathbb{Z}_2)}}{32 z^2 (8 y^2 - 6 z x + x^2)} & 
L_1-\frac{\Lambda_-^{(\mathbb{Z}_2\times \mathbb{Z}_2)}}{32 z^2 (8 y^2 - 6 z x + x^2)} & 
1 \\
\end{pmatrix}
,
\label{eq:LZ2Z2untw}
\end{align}
with 
\begin{align}
L_1\equiv \frac{x^2 y^2 + 8 y (-2 x + y) z^2 - 48 y z^3 + 
 128 z^4}{16 (x - 4 z) (x - 2 z) z^2}.
 \end{align}

We arrive at the physical Yukawa couplings by substituting 
these quantities into Eq. (\ref{eq:physyukawa}). 
In a similar way with analysis on $T^6/\mathbb{Z}_3$ orbifold, we perform the numerical search to 
clarify the structure of physical Yukawa couplings by setting $x=4$. 
Since the value of blow-up mode has to be smaller than the other moduli values to justify our effective action, we choose a relatively large value for $x$, that is, $x=4$. 
We find that physical Yukawa coupling $Y_{\hat{1}\hat{a}\hat{b}}$ becomes 
the rank-two matrix due to the structure of diagonalizing matrix (\ref{eq:LZ2Z2untw}). 
In this respect, the Higgs field is assumed to either elements of $A^2$ or $A^3$, and quarks/leptons are the elements of $(A^1,A^2,A^3)$.

\begin{figure}[H]
 \begin{minipage}{0.49\hsize}
  \begin{center}
   \includegraphics[scale=0.5]{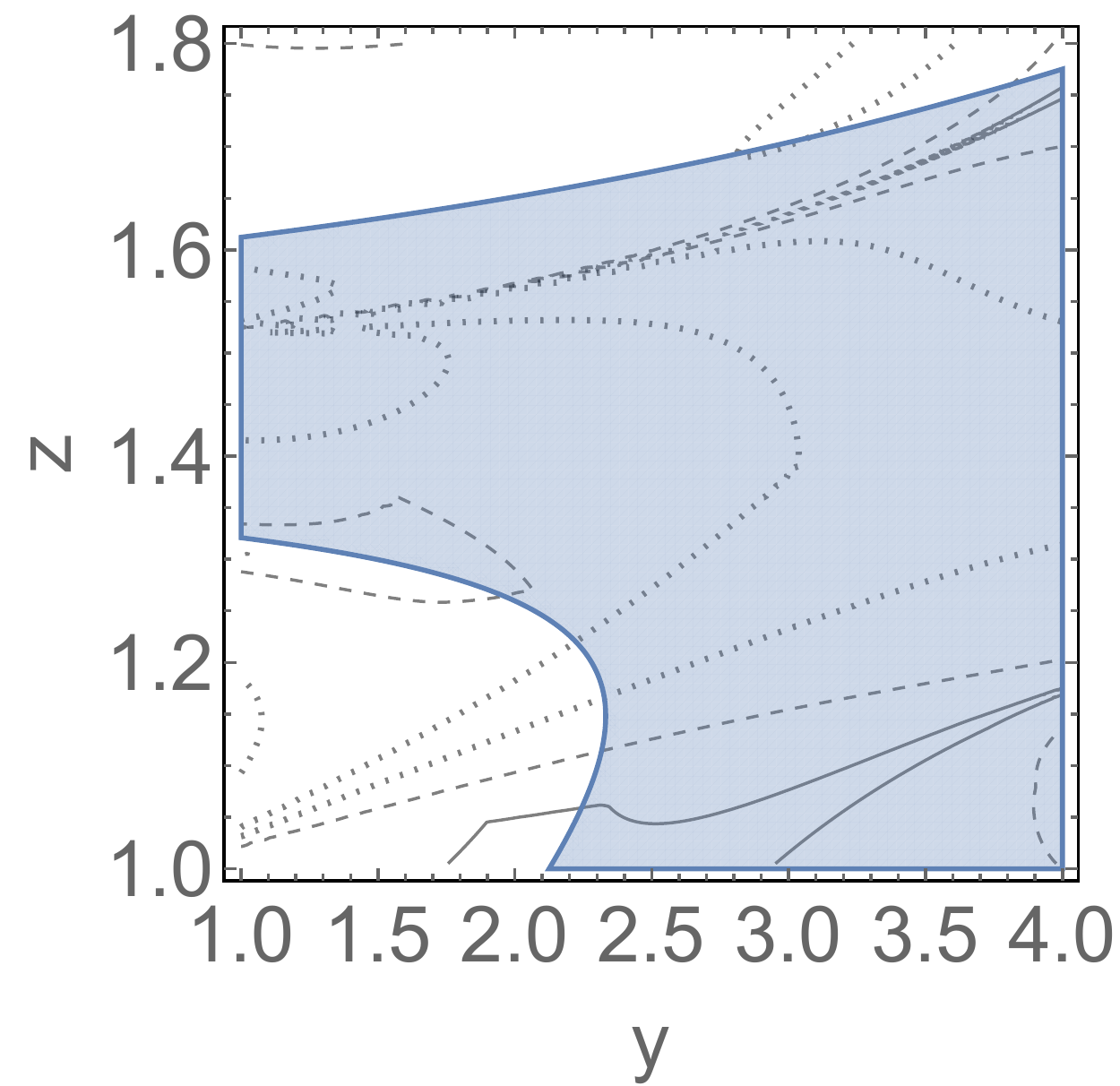}
  \end{center}
 \end{minipage}
 \begin{minipage}{0.49\hsize}
  \begin{center}
    \includegraphics[scale=0.5]{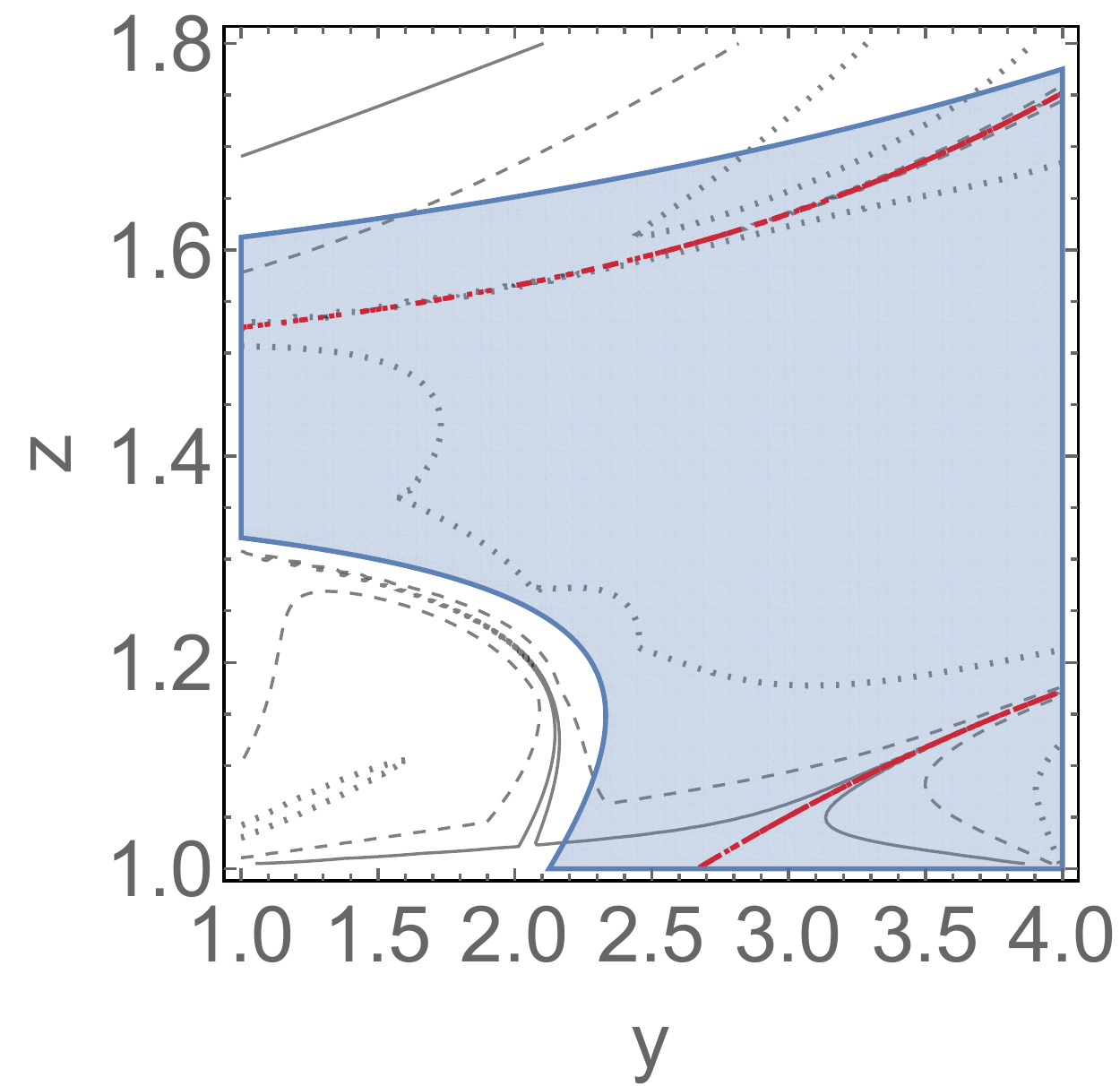}
  \end{center}
 \end{minipage}
\caption{In the left and right panels, we plot two ratios $r_1$ and $r_2$ of eigenvalues of the physical Yukawa couplings ($Y_{\hat{2}\hat{a}\hat{b}}$) to the maximum one with respect to $y$ and $z$, respectively.
The dotted, dashed, and solid curves correspond to the values for $0.5$ $(10^{-1})$, $10^{-1} (10^{-2})$, and $10^{-2} (10^{-3})$, in the left (right) panel. 
Note that Yukawa couplings reduce to the rank-two Yukawa matrix on red curves in the right panel. 
In both panels, the blue shaded regions represent the allowed volume $1<{\cal V}\leq 30$.}
    \label{fig:Z2Z2x42}
\end{figure}

\begin{figure}[H]
 \begin{minipage}{0.49\hsize}
  \begin{center}
    \includegraphics[scale=0.5]{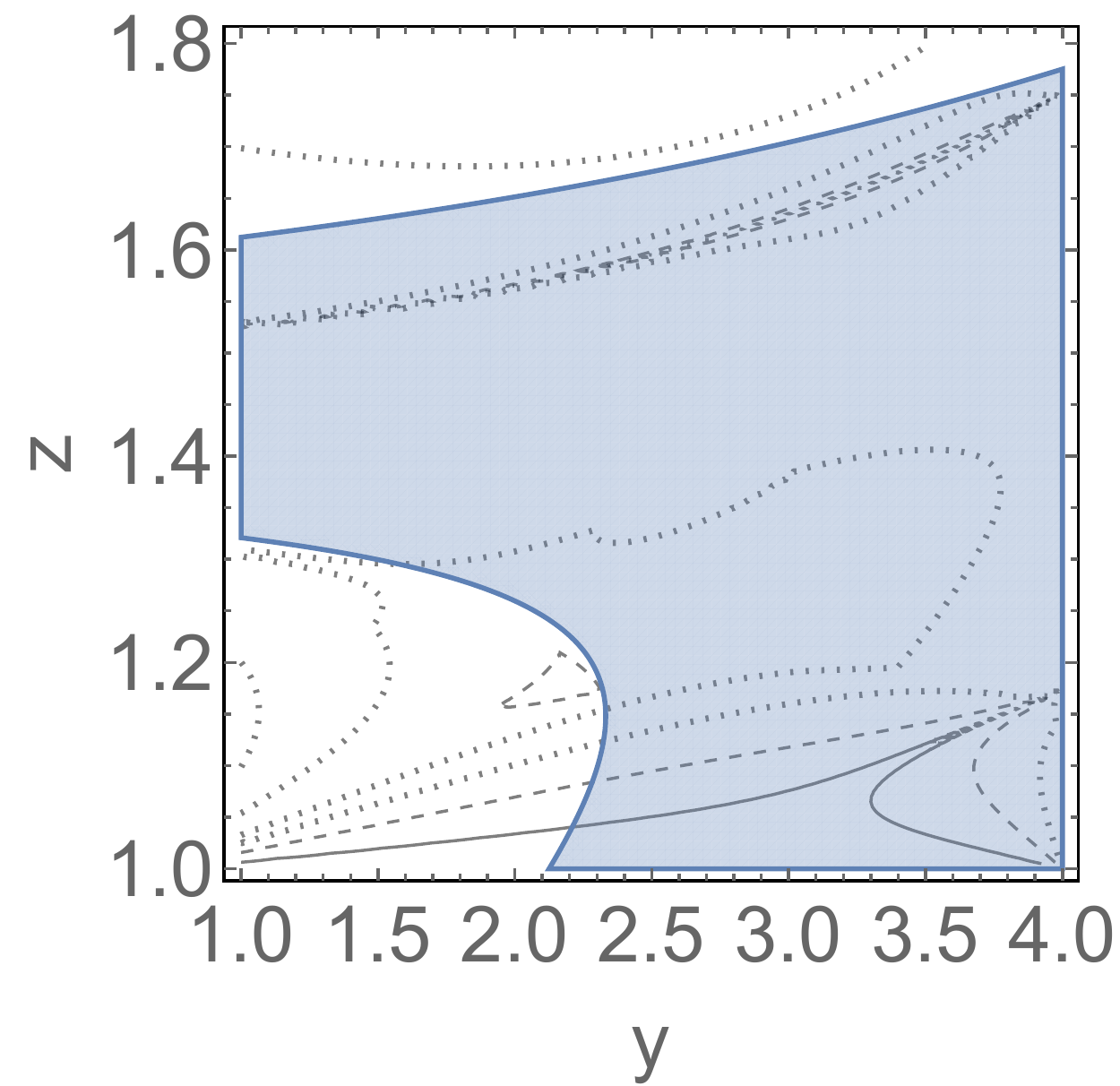}
  \end{center}
 \end{minipage}
 \begin{minipage}{0.49\hsize}
  \begin{center}
    \includegraphics[scale=0.5]{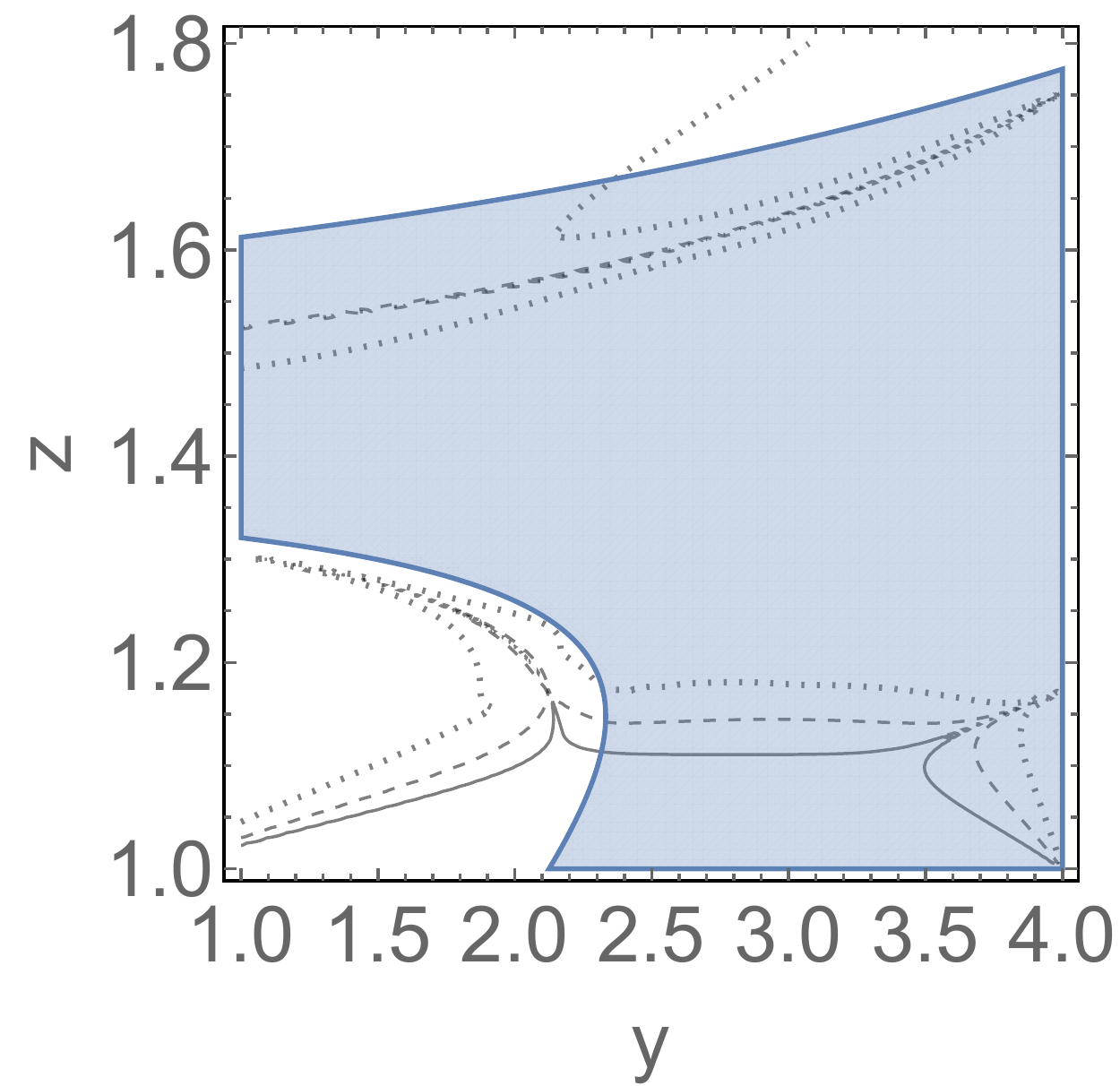}
  \end{center}
 \end{minipage}
\caption{In the left and right panels, we plot two ratios $r_1$ and $r_2$ of eigenvalues of the physical Yukawa couplings ($Y_{\hat{3}\hat{a}\hat{b}}$) to the maximum one with respect to $y$ and $z$, respectively. 
The dotted, dashed, and solid curves correspond to the values for $0.5$ $(10^{-1})$, $10^{-1}(10^{-2})$, and $10^{-2}(10^{-3})$, in the left (right) panel, respectively. 
In both panels, the blue shaded regions represent the allowed volume $1<{\cal V}\leq 30$.}
    \label{fig:Z2Z2x43}
\end{figure}

In Figures \ref{fig:Z2Z2x42} and \ref{fig:Z2Z2x43}, we plot two ratios of eigenvalues of the physical Yukawa couplings $Y_{\hat{2}\hat{a}\hat{b}}$ and $Y_{\hat{3}\hat{a}\hat{b}}$ to the maximum one, 
\begin{align}
    r_1 =\frac{|m_2|}{|m_1|},\quad r_2 =\frac{|m_3|}{|m_1|},
\end{align}
with the order $|m_1|>|m_2|>|m_3|$, respectively. 
The left and right panels in Figures \ref{fig:Z2Z2x42} and \ref{fig:Z2Z2x43} 
correspond to the $r_1$ and $r_2$ with respect 
to $y$ and $z$, respectively. 
In the same way as the results on $T^6/\mathbb{Z}_3$ orbifold, 
${\cal O}(1)$ values of the moduli fields  induce the ${\cal O}(10^{-1})$ and ${\cal O}(10^{-2})$ hierarchical structures 
for $r_1$ and $r_2$, 
because of the existence of blow-up modes appearing in the off-diagonal entries of matter K\"ahler metric. 
Remarkably, there also exists an overlap region between the region with the hierarchical Yukawa 
couplings and the desired value of the toroidal volume ${\cal V}\leq 30$.


%%%%%%%%%%%%%%%%%%%%%%%%%%%%%%%%%%%%%%%%%%%%%%%%%%%%%%%%%%%%
\subsubsection{Yukawa couplings of the twisted mode}
\label{subsubsec:tw}
%%%%%%%%%%%%%%%%%%%%%%%%%%%%%%%%%%%%%%%%%%%%%%%%%%%%%%%%%%%%

In this section, we analyze the Yukawa couplings among two untwisted modes and the twisted mode 
by assuming that the matter fields in the SM are originating from 
the untwisted modes ($A^1,A^2,A^3$) and the twisted mode $A^4$. 
Since the holomorphic Yukawa coupling as well as the K\"ahler metric of the untwisted modes 
($A^1,A^2,A^3$) is totally symmetric, we focus on the Yukawa coupling of ($A^1,A^2,A^4$) without 
loss of generality. 
The holomorphic Yukawa couplings are of the form
\begin{align}
\kappa_{1ab}=\kappa_{2ab}=
\begin{pmatrix}
    0 & 0 & 0 \\
    0 & 0 & 0\\
    0 & 0 & -16 \\
\end{pmatrix}
,
\quad
\kappa_{4ab}=
\begin{pmatrix}
    0 & 0 & -16 \\
    0 & 0 & -16\\
    -16 & -16 & 288 \\
\end{pmatrix}
,
\end{align}
which are rank 1 and 2 matrices, respectively. 
In the following, we examine the structure of physical Yukawa couplings with an emphasis on 
three classes of holomorphic Yukawa couplings, namely (i) $\kappa_{1ab}$, (ii) $\kappa_{2ab}$, (iii) $\kappa_{4ab}$, where the Higgs field is identified with the element of $A^1$, $A^2$ and $A^4$, 
respectively. In all cases, quarks/leptons are assumed to the elements of $(A^1,A^2,A^4)$. 

The moduli K\"ahler metric is calculated as
\scriptsize
\begin{align}
\begin{split}
K_{1\bar{1}}&=\frac{(t^2t^3 - 8(t^4)^2)^2}{4(t^1t^2t^3 -8 (t^1 + t^2 + t^3) (t^4)^2 + 48 (t^4)^3)^{2}},
\quad
K_{2\bar{2}}=\frac{(t^1t^3 - 8(t^4)^2)^2}{4(t^1t^2t^3 -8 (t^1 + t^2 + t^3) (t^4)^2 + 48 (t^4)^3)^{2}},\\
K_{4\bar{4}}&=\frac{16 (t^1 t^2 t^3 (t^1 + t^2 + t^3) - 18 t^1 t^2 t^3 t^4 + 
   8 (t^1 + t^2 + t^3)^2 (t^4)^2 - 96 (t^1 + t^2 + t^3) (t^4)^3 + 432 (t^4)^4)}{4(t^1t^2t^3 -8 (t^1 + t^2 + t^3) (t^4)^2 + 48 (t^4)^3)^{2}},\\
K_{1\bar{2}}&=K_{2\bar{1}}=\frac{8(t^3 - 4t^4)(t^3 - 2t^4) (t^4)^2}{4(t^1t^2t^3 -8 (t^1 + t^2 + t^3) (t^4)^2 + 48 (t^4)^3)^{2}},   
\quad
K_{1\bar{4}}=K_{4\bar{1}}=\frac{16 t^4 (-t^2 t^3 (t^2 + t^3) + 9 t^2 t^3 t^4 - 24 (t^4)^3)}{4(t^1t^2t^3 -8 (t^1 + t^2 + t^3) (t^4)^2 + 48 (t^4)^3)^{2}},\\   
K_{2\bar{4}}&=K_{4\bar{2}}=\frac{16 t^4 (-t^1 t^3 (t^1 + t^3) + 9 t^1 t^3 t^4 - 24 (t^4)^3)}{4(t^1t^2t^3 -8 (t^1 + t^2 + t^3) (t^4)^2 + 48 (t^4)^3)^{2}},
\end{split}
\end{align}
\normalsize
with $t^a={\rm Re}(T^a)$ and hereafter, we adopt the parametrization of the moduli vacuum expectation values (\ref{eq:vev2}) 
in the same way as the previous analysis. 
We omit the complicated expressions for the eigenvalues of the K\"ahler metric as well as its diagonalization matrix. 
The physical Yukawa couplings are calculated from the formula
\begin{align}
Y_{\hat{a}\hat{b}\hat{c}} &= e^{K_{\rm ks}}\sum_{d,e,f=1,2,4} (\Lambda^{-1/2}L)_{\hat{a}}^{d}  (\Lambda^{-1/2}L)_{\hat{b}}^{e} (\Lambda^{-1/2}L)_{\hat{c}}^{f}\kappa_{def}.
\end{align}

\begin{figure}[H]
    \centering
    \includegraphics[scale=0.5]{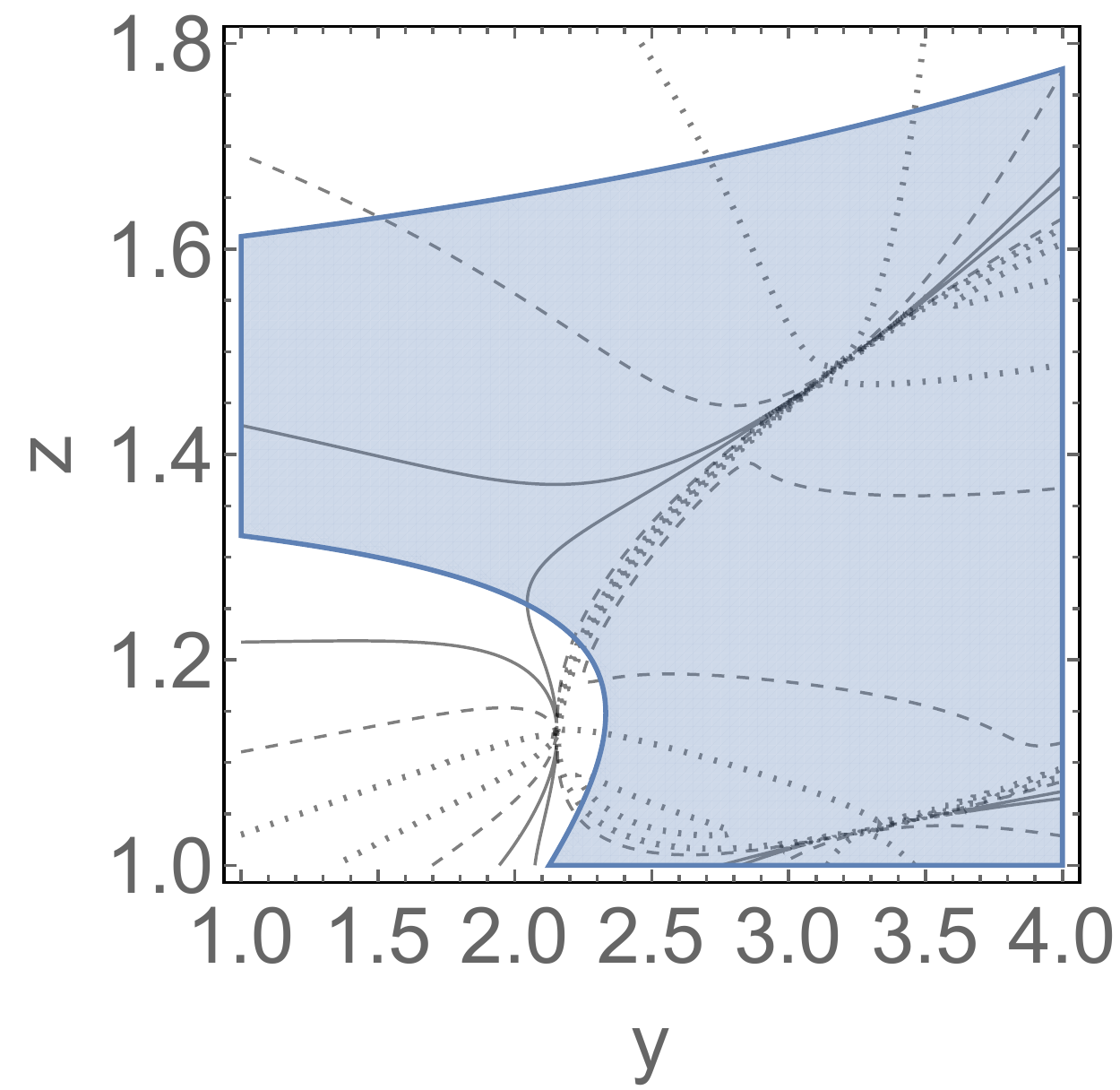}
    \caption{
    We draw the ratio of eigenvalues of the rank-two physical Yukawa couplings ($Y_{\hat{3}\hat{a}\hat{b}}$) to the maximum one with respect to $y$ and $z$. 
The dotted, dashed, and solid curves correspond to the values for $10^{-1}$, $10^{-2}$, and $10^{-3}$, 
and the blue shaded regions represent the allowed volume $1<{\cal V}\leq 30$.}
    \label{fig:Z2Z2twist}
\end{figure}

From our numerical search along the moduli locus (\ref{eq:vev2}), we find that the rank of physical Yukawa couplings in all three cases (i)-(iii) has two 
and there is no significant differences among them. 
Hence, we show the non-vanishing ratio of physical Yukawa coupling in the case (iii) as shown in Figure \ref{fig:Z2Z2twist} with $x=4$. 
The difficulty of realizing the full rank of physical Yukawa couplings would be attributable to the trivial structure of holomorphic Yukawa couplings. 

%%%%%%%%%%%%%%%%%%%%%%%%%%%%%%%%%%%%%%%%%%%%%%%%%%%%%%%%%%%%
\subsection{A mirror dual of $T^6/(\mathbb{Z}_3\times \mathbb{Z}_3)$ orbifold}
\label{subsec:orbifold3}
%%%%%%%%%%%%%%%%%%%%%%%%%%%%%%%%%%%%%%%%%%%%%%%%%%%%%%%%%%%%

As a final example, we deal with a mirror dual of rigid $T^6/(\mathbb{Z}_3\times \mathbb{Z}_3)$ 
orbifold as developed in Ref. \cite{Candelas:1993nd}. 
In contrary to the previous backgrounds, there exist $h^{2,1}=30$ number of complex structure deformation and no K\"ahler deformation. Hence, there will be no constraint on the internal volume 
that determines the value of 4D gauge coupling. 
The chiral zero-modes for our interest are originating from the 
complex structure moduli whose physical Yukawa couplings are analyzed in detail below. 

In this section, we show the  hierarchical structure about the physical Yukawa couplings 
for chiral zero-modes associated with the complex structure moduli.  
In particular, we focus on one of the blow-up modes ($\chi$) among totally 27 blow-up modes, 
which is expanded in the small complex structure regime. 
On the other hand, the bulk complex structure moduli $U^i$ are considered the large 
complex structure regime, that is,
\begin{align}
{\rm Im}(U^i) > 1,\quad |\chi| < 1,
\label{eq:regime}
\end{align}
with $i=1,2,3$. 
Note that we still call ``blow-up modes'' on a mirror dual of $T^6/(\mathbb{Z}_3\times \mathbb{Z}_3)$ 
orbifold, since they correspond to K\"ahler moduli associated with exceptional divisor on a 
rigid $T^6/(\mathbb{Z}_3\times \mathbb{Z}_3)$ orbifold. 
The moduli K\"ahler potential on a mirror dual of rigid $T^6/(\mathbb{Z}_3\times \mathbb{Z}_3)$ 
is known as
\begin{align}
    K=-\ln \Bigl[i(U^1-\widebar{U}^1)(U^2-\widebar{U}^2)(U^3-\widebar{U}^3)\Bigl] 
       + \frac{27\sqrt{3}\Gamma(\frac{1}{3})^6}{(2\pi)^6}\frac{|\chi|^2}{i(U^1 -\widebar{U}^1)(U^2 -\widebar{U}^2)(U^3 -\widebar{U}^3)},
       \label{eq:Kmirror} 
\end{align}
taking into account the regime (\ref{eq:regime}). 
In addition, the holomorphic Yukawa couplings are also found by calculating the 
integrals of the holomorphic three-form over three-cycles. 
The non-vanishing holomorphic Yukawa couplings under the regime (\ref{eq:regime}) 
are given by
\begin{align}
\kappa_{123}=1,%\quad \kappa_{\chi\chi\chi}=27i(2\pi)^{15}e^{\frac{4\pi i}{3} (U^1+U^2+U^3)},
\label{eq:yukawamirror}
\end{align}
where we omit an expression of exponentially-suppressed holomorphic Yukawa couplings of the blow-up mode. 
Since we are interested in assessing to what extent the K\"ahler mixing induces the hierarchical structure of physical 
Yukawa couplings, we focus on the Yukawa couplings of untwisted modes. 
It is remarkable that the blow-up mode is now expanded in the small complex structure regime 
and in the blow-down limit $\chi\rightarrow 0$, other holomorphic Yukawa couplings such as $\kappa_{\chi\chi U^i}$ and $\kappa_{\chi U^iU^j}$ vanish. 
We refer Ref. \cite{Candelas:1993nd} for the derivation of the moduli effective action, which can be achieved by 
calculating the holomorphic three-form of CY threefold embedded into the 
complex projective space $\mathbb{CP}^8$. 

We are ready to analyze the physical Yukawa couplings (\ref{eq:physyukawa}) with the K\"ahler metric given by (\ref{eq:Kmirror}) 
and the holomorphic Yukawa couplings (\ref{eq:yukawamirror}), in particular 
for the untwisted modes. 
The explicit form of the K\"ahler metric is given by
\begin{align}
K_{i\bar{j}}=
    \begin{pmatrix}
       \frac{1}{4(\tau^1)^2} +\frac{27 \sqrt{3} |\chi|^2 \Gamma\left(\frac{1}{3}\right)^6}{1024\pi^6 (\tau^1)^3\tau^2 \tau^3}
       &
       \frac{27 \sqrt{3} |\chi|^2 \Gamma\left(\frac{1}{3}\right)^6}{2048\pi^6 (\tau^1)^2(\tau^2)^2 \tau^3}
       &
       \frac{27 \sqrt{3} |\chi|^2 \Gamma\left(\frac{1}{3}\right)^6}{2048\pi^6 (\tau^1)^2\tau^2 (\tau^3)^2}  
       \\ 
       \frac{27 \sqrt{3} |\chi|^2 \Gamma\left(\frac{1}{3}\right)^6}{2048\pi^6 (\tau^1)^2(\tau^2)^2 \tau^3}
       &
       \frac{1}{4(\tau^2)^2} +\frac{27 \sqrt{3} |\chi|^2 \Gamma\left(\frac{1}{3}\right)^6}{1024\pi^6 \tau^1(\tau^2)^3 \tau^3}
       &       
       \frac{27 \sqrt{3} |\chi|^2 \Gamma\left(\frac{1}{3}\right)^6}{2048\pi^6 \tau^1(\tau^2)^2 (\tau^3)^2}  
       \\ 
       \frac{27 \sqrt{3} |\chi|^2 \Gamma\left(\frac{1}{3}\right)^6}{2048\pi^6 (\tau^1)^2\tau^2 (\tau^3)^2}  
       &       
       \frac{27 \sqrt{3} |\chi|^2 \Gamma\left(\frac{1}{3}\right)^6}{2048\pi^6 \tau^1(\tau^2)^2 (\tau^3)^2}  
       &
       \frac{1}{4(\tau^3)^2} +\frac{27 \sqrt{3} |\chi|^2 \Gamma\left(\frac{1}{3}\right)^6}{1024\pi^6 \tau^1\tau^2 (\tau^3)^3}
    \end{pmatrix}
    \quad (i,j=1,2,3),
    \label{eq:Kijmirror}
\end{align}
with $\tau^i \equiv {\rm Im}(U^i)$, 
where the value of blow-up mode $|\chi|$ encodes the amount of kinetic mixing associated 
to a non-diagonal K\"ahler metric. 
In the blow-down limit $|\chi|\rightarrow 0$, the off-diagonal components of the K\"ahler metric 
go to zero, and these play an important role 
of realizing the rank-three physical Yukawa matrix with the hierarchical structure as shown below. 

\begin{figure}[H]
 \begin{minipage}{0.49\hsize}
  \begin{center}
    \includegraphics[scale=0.5]{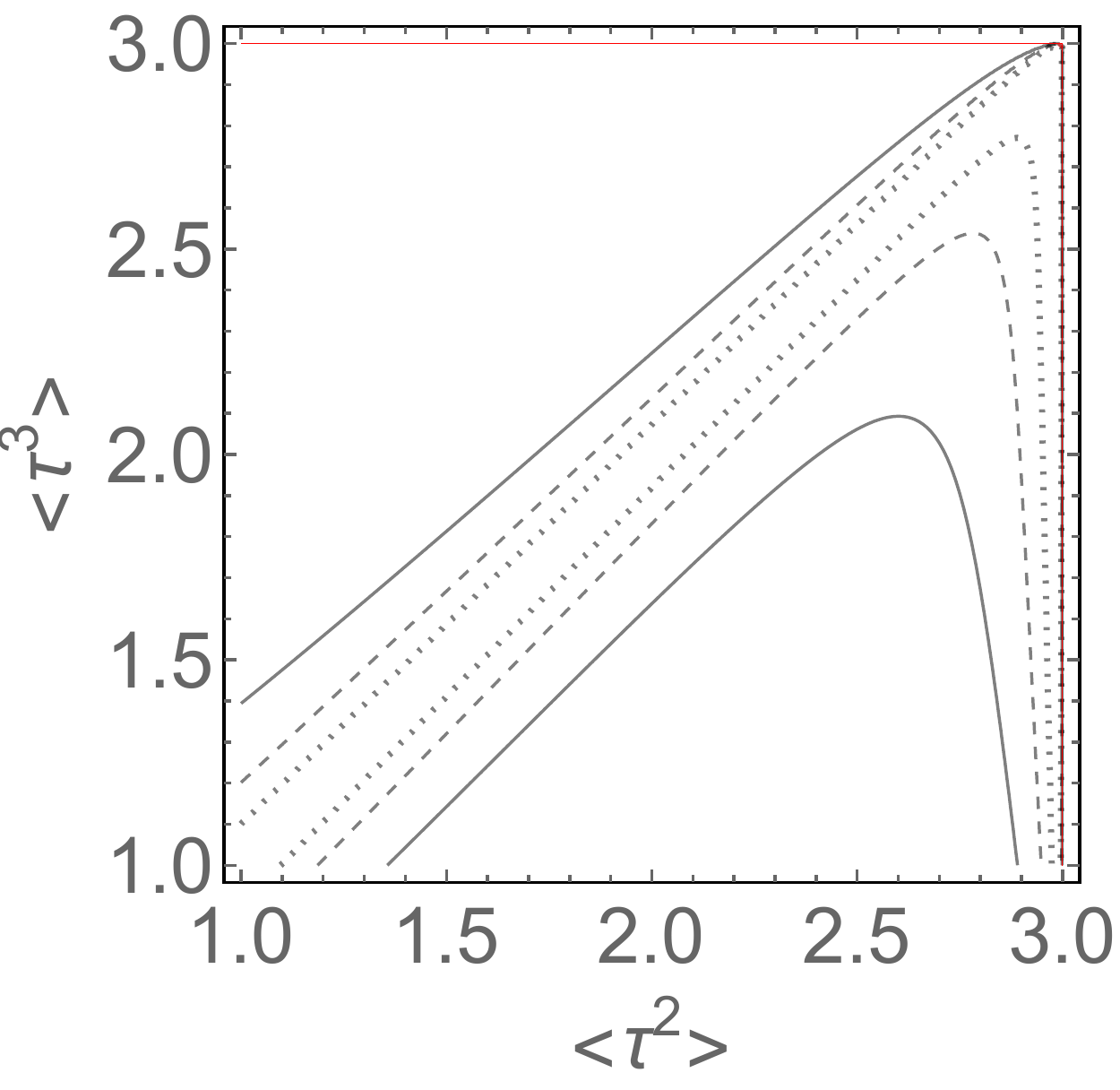}
  \end{center}
 \end{minipage}
 \begin{minipage}{0.49\hsize}
  \begin{center}
    \includegraphics[scale=0.5]{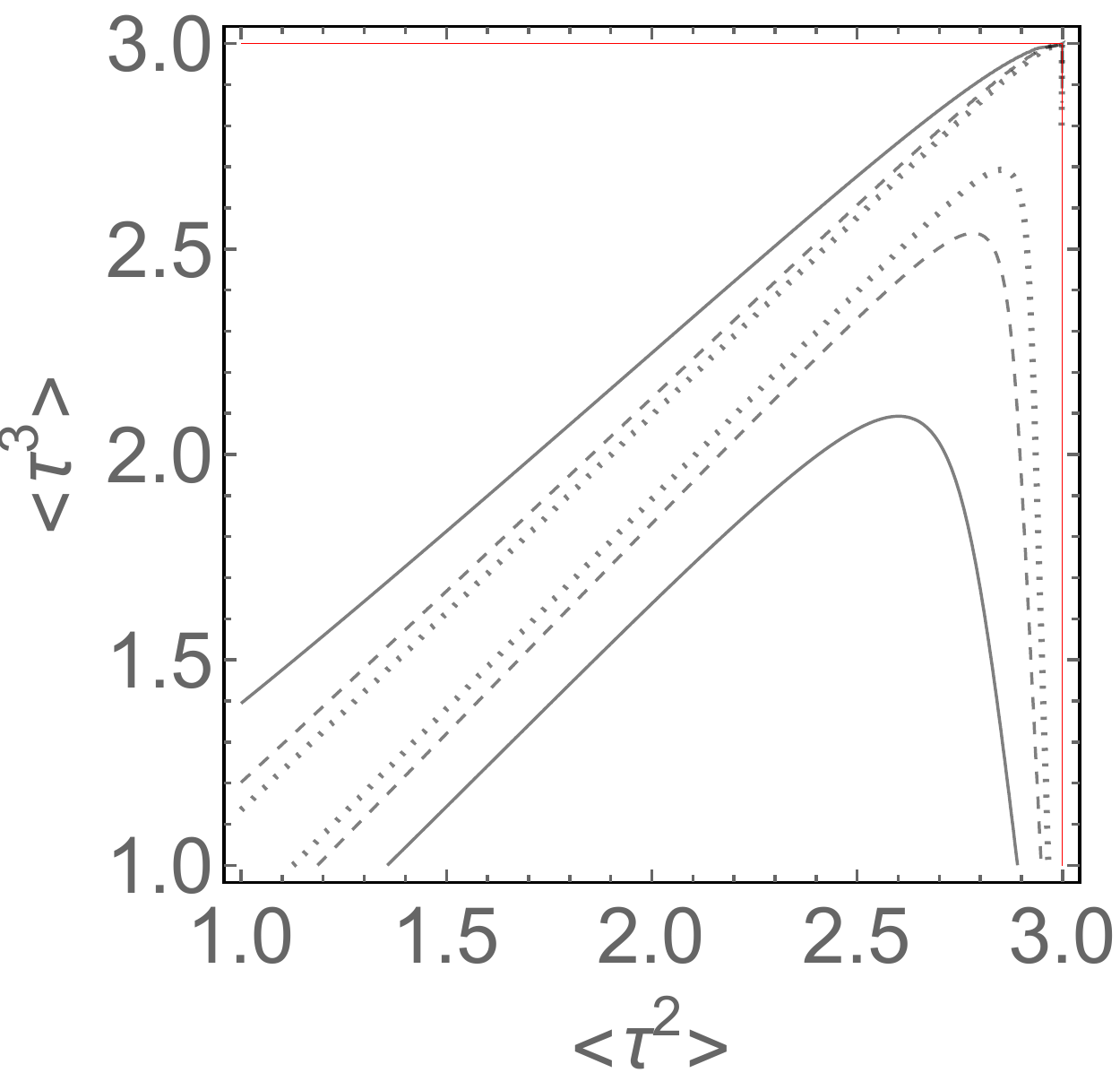}
  \end{center}
 \end{minipage}
\caption{We plot the smallest ratio $r_2$ of eigenvalues of the physical Yukawa couplings 
$Y_{\hat{3}\hat{a}\hat{b}}$ to the maximum one with respect to $\langle\tau^2\rangle$ and $\langle \tau^3\rangle$ by setting $|\chi|=0.1$ and $|\chi|=0.01$ in the left and right panels, respectively and other moduli values are set as in Eq. (\ref{eq:modulivevsmirror}). 
In both panels, red lines correspond to $r_2=0$, namely rank-two Yukawa matrix. 
In the left (right) panel, the dotted, dashed and solid curves correspond to the values for $4\times 10^{-5} (3\times 10^{-7})$, $2\times 10^{-5} (2\times 10^{-7})$, and $10^{-5} (10^{-7})$, respectively.}
    \label{fig:mirror}
\end{figure}

We numerically analyze the structure of physical Yukawa couplings $Y_{\hat{1}\hat{i}\hat{j}}$, $Y_{\hat{2}\hat{i}\hat{j}}$ and $Y_{\hat{3}\hat{i}\hat{j}}$ 
by setting 
\begin{align}
    {\rm Re}(U^1) = {\rm Re}(U^2) = {\rm Re}(U^3) = 0,\quad
    \tau^1 = 3,
\label{eq:modulivevsmirror}
\end{align}
and the other moduli values ($\tau^2$,$\tau^3$,$\chi$) are chosen as free 
parameters. 
It turns out that the rank of physical Yukawa couplings  $\{Y_{\hat{1}\hat{i}\hat{j}},Y_{\hat{2}\hat{i}\hat{j}},Y_{\hat{3}\hat{i}\hat{j}}\}$ is three in the moduli space except 
for $\tau^i=\tau^j$ with $i\neq j$. 
When at least two moduli values are coincides with each other, two eigenvalues of the 
K\"ahler metric are degenerate due to the symmetric structure of K\"ahler metric 
with respect to $\tau^i$. 
The phenomenon is observed in the following numerical results. 
In a general moduli space, eigenvalues of physical Yukawa 
couplings $(m_1,m_2,m_3)$ with $|m_1|\geq |m_2|\geq |m_3|$ satisfy the relation
\begin{align}
|m_2| \simeq |m_1|,
\end{align}
for all $\{Y_{\hat{1}\hat{i}\hat{j}},Y_{\hat{2}\hat{i}\hat{j}},Y_{\hat{3}\hat{i}\hat{j}}\}$. 
Since we focus on the moduli space within (\ref{eq:regime}), 
the off-diagonal entries in K\"ahler metric (\ref{eq:Kijmirror}) should be smaller than the 
diagonal one. 
It results in two degenerate eigenvalues of physical Yukawa couplings and those are determined by the rank-two holomorphic Yukawa matrix (\ref{eq:holeg}). 
The smallest eigenvalue of physical Yukawa coupling has an origin in the smallness 
of the off-diagonal entries in the K\"ahler metric. 
Indeed, the ratio $r_2=|m_3|/|m_1|$ is sensitive to the value of the blow-up modulus $\chi$ as 
indicated in Figure \ref{fig:mirror}. 
We plot the ratio $r_2$ of the physical Yukawa coupling $Y_{\hat{3}\hat{i}\hat{j}}$, but the same structure 
arises in the other physical Yukawa couplings. 

It is remarkable that there exist a huge hierarchy between $m_3$ and others, due to 
the suppressed off-diagonal entries in K\"ahler metric (\ref{eq:Kijmirror}). 
Indeed, the ratio of off-diagonal components to diagonal one in $|\chi|\rightarrow 0$ limit 
is typically of ${\cal O}(3.5\times 10^{-2}|\chi|^2(\tau^i)^{-3})$, when the moduli values $\tau^i$ are assumed to be the same order. 
From the formula of physical Yukawa couplings (\ref{eq:physyukawa}), the smallest ratio of eigenvalues of 
the physical Yukawa couplings, $r_2$, is then further suppressed by the value of blow-up mode through the 
matter K\"ahler metric. 
That phenomenon is consistent with our numerical results, indicating that the ratio $r_2$ becomes smaller and smaller as $|\chi|$ decreases.

%%%%%%%%%%%%%%%%%%%%%%%%%%%%%%%%%%%%%%%%%%%%%%%%%%%%%%%%%%%%
%%%%%%%%%%%%%%%%%%%%%%%%%%%%%%%%%%%%%%%%%%%%%%%%%%%%%%%%%%%%
\section{Conclusions}
\label{sec:con}
%%%%%%%%%%%%%%%%%%%%%%%%%%%%%%%%%%%%%%%%%%%%%%%%%%%%%%%%%%%%
%%%%%%%%%%%%%%%%%%%%%%%%%%%%%%%%%%%%%%%%%%%%%%%%%%%%%%%%%%%%

The non-trivial structure of physical Yukawa couplings has an origin 
in the matter field K\"ahler metric and the 
holomorphic Yukawa couplings in the context of 4D ${\cal N}=1$ 
supersymmetric theory. 
In the context of heterotic orbifold models, it was known that hierarchical 
Yukawa couplings are realized in the twisted sector due to the 
exponentially-suppressed holomorphic Yukawa couplings, whereas 
the Yukawa couplings among the untwisted modes are typically of ${\cal O}(1)$ in the heterotic 
string theory with standard embedding, where matter fields 
are assigned to the fundamental or anti-fundamental representations 
of $E_6$ gauge group. 
Hence, we have mainly focused on the Yukawa couplings of untwisted modes 
whether the matter field K\"ahler metric provide their hierarchical structure. 
Since the matter fields $27$ and $\widebar{27}$ are respectively in one-to-one correspondence 
with the K\"ahler moduli and complex structure moduli, one can 
argue definite properties of the matter field K\"ahler metric in the physical 
Yukawa couplings.

We have analyzed $27^3$ and $\widebar{27}^3$ Yukawa couplings of untwisted modes on a 
simplified CY threefold in Section \ref{subsubsec:CY} and the untwisted and twisted modes on 
toroidal orbifolds preserving the supersymmetry in Section \ref{subsubsec:orbifolds}. 
It turned out that off-diagonal entries in matter field K\"ahler metric 
are necessary to generate the rank-full physical Yukawa couplings with 
hierarchical structure. 
There exists the non-vanishing matter K\"ahler mixing on unfactorizable tori such 
as $T^6/\mathbb{Z}_{3,4,6}$ orbifolds and blown-up toroidal 
orbifolds. 
Indeed, we performed the numerical search on $T^6/\mathbb{Z}_3$ 
orbifold, blown-up $T^6/(\mathbb{Z}_2\times \mathbb{Z}_2)$ orbifold, 
and a mirror dual of $T^6/(\mathbb{Z}_3\times \mathbb{Z}_3)$ 
orbifold. 
In the former two examples, we exhibited that the hierarchical structure of 
physical Yukawa couplings of matter zero-modes associated with K\"ahler moduli 
is provided by ${\cal O}(1)$ values 
of moduli fields, taking into account the realistic value of 4D gauge couplings. 
In the latter example, the complex structure moduli space accompanying 
chiral matter $\widebar{27}$ was analyzed in detail. 
We found that the small but non-zero value of the blow-up mode, which  is required to justify the 
effective action, leads to the suppressed off-diagonal components of the 
matter field K\"ahler metric, thereby providing the hierarchical structure 
of physical $\widebar{27}^3$ Yukawa couplings of untwisted modes.

So far, we have focused on the realization of hierarchical structure of 
physical Yukawa couplings of untwisted modes by means of the structure of matter 
K\"ahler metric. 
These scenarios would be widely applicable to more realistic models such as 
$SU(5)$ or $SO(10)$ grand unified theories achieved by introducing Wilson-lines 
on the Cartan directions of $E_6$ and theories with smaller gauge groups. 
For example, many realistic models have been constructed in 
$T^6/\mathbb{Z}_{6-II}$ orbifold models \cite{Kobayashi:2004ya,Buchmuller:2006ik,Lebedev:2006kn}.
In these models, the third generation of left-handed (right-handed) quarks and leptons 
corresponds to the untwisted sector $A^1$ ($A^2$) and the Higgs field  
corresponds to the untwisted mode $A^3$, while the other two generations of 
quarks and leptons are originated from twisted sectors.
At the orbifold limit, the Yukawa matrices are rank-one, and 
only the third generation can gain masses  \cite{Ko:2007dz}.
Thus, it would be interesting to extend our analysis to above models 
on blown-up  $T^6/\mathbb{Z}_{6-II}$ orbifold \footnote{
See for an approach of blown-up $T^6/\mathbb{Z}_{6-II}$ orbifold,  Ref.~\cite{Nibbelink:2009sp}.}.
Including these topics, 
we leave a pursue of more phenomenologically interesting three-generation models for future work.

The moduli vacuum expectation values are 
assumed to be free parameters in our analysis to find the structure of physical Yukawa couplings. 
However, a realization of the hierarchical structure of physical Yukawa couplings is 
highly dependent on the stabilization mechanism of the moduli fields. 
The moduli stabilization in the context of heterotic string theory is 
still open challenging issue, since it generically induces the backreaction on the 
background geometry. 
We comment on the possible mechanism of the moduli stabilization which we leave for 
future work. 
When the matter zero-modes have an origin in the K\"ahler moduli sector, 
we require some non-perturbative effects to stabilize the K\"ahler moduli. 
We often employ the gaugino condensation on the hidden sector with 
one-loop threshold corrections depending on the K\"ahler moduli and/or 
world-sheet instanton effects on the two-cycles of CY threefolds. 
When the matter zero-modes have an origin in the complex structure moduli sector, 
a flux compactification is a powerful tool to determine the vacuum expectation 
value of these moduli fields. Indeed, the flux compactification on a mirror dual of rigid 
CY threefolds in the example of Section \ref{subsec:orbifold3} is suffice to stabilize all the 
moduli fields. We hope to come back to these problems in the future.

%%%%%%%%%%%%%%%%%%%%%%%%%%%%%%%%%%%%%%%%%%%%%%%%%%%%%%%%%%%%
%%%%%%%%%%%%%%%%%%%%%%%%%%%%%%%%%%%%%%%%%%%%%%%%%%%%%%%%%%%%
\subsection*{Acknowledgements}
%%%%%%%%%%%%%%%%%%%%%%%%%%%%%%%%%%%%%%%%%%%%%%%%%%%%%%%%%%%%
%%%%%%%%%%%%%%%%%%%%%%%%%%%%%%%%%%%%%%%%%%%%%%%%%%%%%%%%%%%%

T. K. was supported in part by MEXT KAKENHI Grant Number JP19H04605. H. O. was
supported in part by JSPS KAKENHI Grant Numbers JP19J00664 and JP20K14477.

%%%%%


\begin{thebibliography}{99}

%\cite{ArkaniHamed:1999dc}
\bibitem{ArkaniHamed:1999dc}
N.~Arkani-Hamed and M.~Schmaltz,
%``Hierarchies without symmetries from extra dimensions,''
Phys. Rev. D \textbf{61} (2000), 033005
%doi:10.1103/PhysRevD.61.033005
[arXiv:hep-ph/9903417 [hep-ph]].


%\cite{Kaplan:2000av}
\bibitem{Kaplan:2000av}
D.~E.~Kaplan and T.~M.~P.~Tait,
%``Supersymmetry breaking, fermion masses and a small extra dimension,''
JHEP \textbf{06} (2000), 020
%doi:10.1088/1126-6708/2000/06/020
[arXiv:hep-ph/0004200 [hep-ph]].


%\cite{Froggatt:1978nt}
\bibitem{Froggatt:1978nt}
C.~D.~Froggatt and H.~B.~Nielsen,
%``Hierarchy of Quark Masses, Cabibbo Angles and CP Violation,''
Nucl. Phys. B \textbf{147}, 277-298 (1979).
%doi:10.1016/0550-3213(79)90316-X

%\cite{Ishimori:2010au}
\bibitem{Ishimori:2010au}
H.~Ishimori, T.~Kobayashi, H.~Ohki, Y.~Shimizu, H.~Okada and M.~Tanimoto,
%``Non-Abelian Discrete Symmetries in Particle Physics,''
Prog. Theor. Phys. Suppl. \textbf{183}, 1-163 (2010)
%doi:10.1143/PTPS.183.1
[arXiv:1003.3552 [hep-th]];
%
%\cite{Ishimori:2012zz}
%\bibitem{Ishimori:2012zz}
%H.~Ishimori, T.~Kobayashi, H.~Ohki, H.~Okada, Y.~Shimizu and M.~Tanimoto,
%``An introduction to non-Abelian discrete symmetries for particle physicists,''
Lect. Notes Phys. \textbf{858}, 1-227 (2012).
%doi:10.1007/978-3-642-30805-5


%\cite{Hamidi:1986vh}
\bibitem{Hamidi:1986vh}
S.~Hamidi and C.~Vafa,
%``Interactions on Orbifolds,''
Nucl. Phys. B \textbf{279} (1987), 465-513.
%doi:10.1016/0550-3213(87)90006-X


%\cite{Dixon:1986qv}
\bibitem{Dixon:1986qv}
L.~J.~Dixon, D.~Friedan, E.~J.~Martinec and S.~H.~Shenker,
%``The Conformal Field Theory of Orbifolds,''
Nucl. Phys. B \textbf{282} (1987), 13-73.
%doi:10.1016/0550-3213(87)90676-6


%\cite{Burwick:1990tu}
\bibitem{Burwick:1990tu}
T.~T.~Burwick, R.~K.~Kaiser and H.~F.~Muller,
%``General Yukawa couplings of strings on Z(N) orbifolds,''
Nucl. Phys. B \textbf{355} (1991), 689-711.
%doi:10.1016/0550-3213(91)90491-F


%\cite{Kobayashi:1991rp}
\bibitem{Kobayashi:1991rp}
T.~Kobayashi and N.~Ohtsubo,
%``Geometrical aspects of Z(N) orbifold phenomenology,''
Int. J. Mod. Phys. A \textbf{9}, 87-126 (1994).
%doi:10.1142/S0217751X94000054

%\cite{Casas:1991ac}
\bibitem{Casas:1991ac}
J.~A.~Casas, F.~Gomez and C.~Munoz,
%``Complete structure of Z(n) Yukawa couplings,''
Int. J. Mod. Phys. A \textbf{8}, 455-506 (1993)
%doi:10.1142/S0217751X93000187
[arXiv:hep-th/9110060 [hep-th]].


%\cite{Ko:2004ic}
\bibitem{Ko:2004ic}
P.~Ko, T.~Kobayashi and J.~h.~Park,
%``Quark masses and mixing angles in heterotic orbifold models,''
Phys. Lett. B \textbf{598}, 263-272 (2004)
%doi:10.1016/j.physletb.2004.08.007
[arXiv:hep-ph/0406041 [hep-ph]].

%\cite{Ko:2005sh}
\bibitem{Ko:2005sh}
P.~Ko, T.~Kobayashi and J.~h.~Park,
%``Lepton masses and mixing angles from heterotic orbifold models,''
Phys. Rev. D \textbf{71}, 095010 (2005)
%doi:10.1103/PhysRevD.71.095010
[arXiv:hep-ph/0503029 [hep-ph]].



%\cite{Cremades:2004wa}
\bibitem{Cremades:2004wa}
D.~Cremades, L.~E.~Ibanez and F.~Marchesano,
%``Computing Yukawa couplings from magnetized extra dimensions,''
JHEP \textbf{05} (2004), 079
%doi:10.1088/1126-6708/2004/05/079
[arXiv:hep-th/0404229 [hep-th]].



%\cite{Abe:2008fi}
\bibitem{Abe:2008fi}
H.~Abe, T.~Kobayashi and H.~Ohki,
%``Magnetized orbifold models,''
JHEP \textbf{09} (2008), 043
%doi:10.1088/1126-6708/2008/09/043
[arXiv:0806.4748 [hep-th]].


%\cite{Kobayashi:2019fma}
\bibitem{Kobayashi:2019fma}
T.~Kobayashi, H.~Otsuka and H.~Uchida,
%``Wavefunctions and Yukawa couplings on resolutions of T$^{2}$/\ensuremath{\mathbb{Z}}$_{N}$ orbifolds,''
JHEP \textbf{08} (2019), 046
%doi:10.1007/JHEP08(2019)046
[arXiv:1904.02867 [hep-th]].


%\cite{Abe:2015bxa}
\bibitem{Abe:2015bxa}
H.~Abe, A.~Oikawa and H.~Otsuka,
%``Wavefunctions on magnetized branes in the conifold,''
JHEP \textbf{07} (2016), 054
%doi:10.1007/JHEP07(2016)054
[arXiv:1510.03407 [hep-th]].


%\cite{Blesneag:2018ygh}
\bibitem{Blesneag:2018ygh}
\c{S}.~Blesneag, E.~I.~Buchbinder, A.~Constantin, A.~Lukas and E.~Palti,
%``Matter field K\"ahler metric in heterotic string theory from localisation,''
JHEP \textbf{04} (2018), 139
%doi:10.1007/JHEP04(2018)139
[arXiv:1801.09645 [hep-th]].


%\cite{Dixon:1989fj}
\bibitem{Dixon:1989fj}
L.~J.~Dixon, V.~Kaplunovsky and J.~Louis,
%``On Effective Field Theories Describing (2,2) Vacua of the Heterotic String,''
Nucl. Phys. B \textbf{329} (1990), 27-82.
%doi:10.1016/0550-3213(90)90057-K


%\cite{Green:1987mn}
\bibitem{Green:1987mn}
M.~B.~Green, J.~H.~Schwarz and E.~Witten,
``SUPERSTRING THEORY. VOL. 2: LOOP AMPLITUDES, ANOMALIES AND PHENOMENOLOGY,'' 7 1988.
%257 citations counted in INSPIRE as of 13 Mar 2021


%\cite{Polchinski:1998rr}
\bibitem{Polchinski:1998rr}
J.~Polchinski,
``String theory. Vol. 2: Superstring theory and beyond,'' 
Cambridge Monographson Mathematical Physics, Cambridge University Press, 12 2007.
%doi:10.1017/CBO9780511618123


%\cite{Strominger:1985it}
\bibitem{Strominger:1985it}
A.~Strominger and E.~Witten,
%``New Manifolds for Superstring Compactification,''
Commun. Math. Phys. \textbf{101} (1985), 341.
%doi:10.1007/BF01216094


%\cite{Candelas:1987se}
\bibitem{Candelas:1987se}
P.~Candelas,
%``Yukawa Couplings Between (2,1) Forms,''
Nucl. Phys. B \textbf{298} (1988), 458.
%doi:10.1016/0550-3213(88)90351-3


%\cite{Kreuzer:2000xy}
\bibitem{Kreuzer:2000xy}
M.~Kreuzer and H.~Skarke,
%``Complete classification of reflexive polyhedra in four-dimensions,''
Adv. Theor. Math. Phys. \textbf{4} (2002), 1209-1230
%doi:10.4310/ATMP.2000.v4.n6.a2
[arXiv:hep-th/0002240 [hep-th]].


%\cite{Kreuzerdataset}
\bibitem{Kreuzerdataset}
M.~Kreuzer and H.~Skarke, "http://hep.itp.tuwien.ac.at/ kreuzer/CY/".


%\cite{Candelas:2008wb}
\bibitem{Candelas:2008wb}
P.~Candelas and R.~Davies,
%``New Calabi-Yau Manifolds with Small Hodge Numbers,''
Fortsch. Phys. \textbf{58} (2010), 383-466
%doi:10.1002/prop.200900105
[arXiv:0809.4681 [hep-th]].


%\cite{Greene:1986bm}
\bibitem{Greene:1986bm}
B.~R.~Greene, K.~H.~Kirklin, P.~J.~Miron and G.~G.~Ross,
%``A Three Generation Superstring Model. 1. Compactification and Discrete Symmetries,''
Nucl. Phys. B \textbf{278} (1986), 667-693.
%doi:10.1016/0550-3213(86)90057-X


%\cite{Greene:1986jb}
\bibitem{Greene:1986jb}
B.~R.~Greene, K.~H.~Kirklin, P.~J.~Miron and G.~G.~Ross,
%``A Three Generation Superstring Model. 2. Symmetry Breaking and the Low-Energy Theory,''
Nucl. Phys. B \textbf{292} (1987), 606-652.
%doi:10.1016/0550-3213(87)90662-6



%\cite{Dixon:1986jc}
\bibitem{Dixon:1986jc}
L.~J.~Dixon, J.~A.~Harvey, C.~Vafa and E.~Witten,
%``Strings on Orbifolds. 2.,''
Nucl. Phys. B \textbf{274}, 285-314 (1986).
%doi:10.1016/0550-3213(86)90287-7



%\cite{Ibanez:1987pj}
\bibitem{Ibanez:1987pj}
L.~E.~Ibanez, J.~Mas, H.~P.~Nilles and F.~Quevedo,
%``Heterotic Strings in Symmetric and Asymmetric Orbifold Backgrounds,''
Nucl. Phys. B \textbf{301} (1988), 157-196.
%doi:10.1016/0550-3213(88)90166-6


%\cite{Font:1988mk}
\bibitem{Font:1988mk}
A.~Font, L.~E.~Ibanez and F.~Quevedo,
%``$Z(N$) X $Z$(m) Orbifolds and Discrete Torsion,''
Phys. Lett. B \textbf{217} (1989), 272-276.
%doi:10.1016/0370-2693(89)90864-2

%\cite{Katsuki:1989bf}
\bibitem{Katsuki:1989bf}
Y.~Katsuki, Y.~Kawamura, T.~Kobayashi, N.~Ohtsubo, Y.~Ono and K.~Tanioka,
%``Z(N) ORBIFOLD MODELS,''
Nucl. Phys. B \textbf{341}, 611-640 (1990).
%doi:10.1016/0550-3213(90)90542-L


%\cite{Fischer:2013qza}
\bibitem{Fischer:2013qza}
M.~Fischer, S.~Ramos-Sanchez and P.~K.~S.~Vaudrevange,
%``Heterotic non-Abelian orbifolds,''
JHEP \textbf{07} (2013), 080
%doi:10.1007/JHEP07(2013)080
[arXiv:1304.7742 [hep-th]].


%\cite{Ferrara:1986qn}
\bibitem{Ferrara:1986qn}
S.~Ferrara, C.~Kounnas and M.~Porrati,
%``General Dimensional Reduction of Ten-Dimensional Supergravity and Superstring,''
Phys. Lett. B \textbf{181} (1986), 263.
%doi:10.1016/0370-2693(86)90043-2


%\cite{Cvetic:1988yw}
\bibitem{Cvetic:1988yw}
M.~Cvetic, J.~Louis and B.~A.~Ovrut,
%``A String Calculation of the Kahler Potentials for Moduli of Z(N) Orbifolds,''
Phys. Lett. B \textbf{206} (1988), 227-233.
%doi:10.1016/0370-2693(88)91497-9


%\cite{Ibanez:1992hc}
\bibitem{Ibanez:1992hc}
L.~E.~Ibanez and D.~Lust,
%``Duality anomaly cancellation, minimal string unification and the effective low-energy Lagrangian of 4-D strings,''
Nucl. Phys. B \textbf{382} (1992), 305-361
%doi:10.1016/0550-3213(92)90189-I
[arXiv:hep-th/9202046 [hep-th]].


%\cite{Abe:2015xua}
\bibitem{Abe:2015xua}
H.~Abe, T.~Kobayashi, H.~Otsuka, Y.~Takano and T.~H.~Tatsuishi,
%``Gauge coupling unification in SO(32) heterotic string theory with magnetic fluxes,''
PTEP \textbf{2016} (2016) no.5, 053B01
%doi:10.1093/ptep/ptw038
[arXiv:1507.04127 [hep-ph]].


%\cite{Cvetic:1989ii}
\bibitem{Cvetic:1989ii}
M.~Cvetic, B.~A.~Ovrut and J.~Louis,
%``The Zamolodchikov Metric and Effective Lagrangians in String Theory,''
Phys. Rev. D \textbf{40} (1989), 684.
%doi:10.1103/PhysRevD.40.684


%\cite{Denef:2005mm}
\bibitem{Denef:2005mm}
F.~Denef, M.~R.~Douglas, B.~Florea, A.~Grassi and S.~Kachru,
%``Fixing all moduli in a simple f-theory compactification,''
Adv. Theor. Math. Phys. \textbf{9} (2005) no.6, 861-929
%doi:10.4310/ATMP.2005.v9.n6.a1
[arXiv:hep-th/0503124 [hep-th]].


%\cite{Blaszczyk:2010db}
\bibitem{Blaszczyk:2010db}
M.~Blaszczyk, S.~Groot Nibbelink, F.~Ruehle, M.~Trapletti and P.~K.~S.~Vaudrevange,
%``Heterotic MSSM on a Resolved Orbifold,''
JHEP \textbf{09} (2010), 065
%doi:10.1007/JHEP09(2010)065
[arXiv:1007.0203 [hep-th]].


%\cite{Candelas:1993nd}
\bibitem{Candelas:1993nd}
P.~Candelas, E.~Derrick and L.~Parkes,
%``Generalized Calabi-Yau manifolds and the mirror of a rigid manifold,''
Nucl. Phys. B \textbf{407} (1993), 115-154
%doi:10.1016/0550-3213(93)90276-U
[arXiv:hep-th/9304045 [hep-th]].

%\cite{Kobayashi:2004ya}
\bibitem{Kobayashi:2004ya}
T.~Kobayashi, S.~Raby and R.~J.~Zhang,
%``Searching for realistic 4d string models with a Pati-Salam symmetry: Orbifold grand unified theories from heterotic string compactification on a Z(6) orbifold,''
Nucl. Phys. B \textbf{704}, 3-55 (2005)
%doi:10.1016/j.nuclphysb.2004.10.035
[arXiv:hep-ph/0409098 [hep-ph]].

%\cite{Buchmuller:2006ik}
\bibitem{Buchmuller:2006ik}
W.~Buchmuller, K.~Hamaguchi, O.~Lebedev and M.~Ratz,
%``Supersymmetric Standard Model from the Heterotic String (II),''
Nucl. Phys. B \textbf{785}, 149-209 (2007)
%doi:10.1016/j.nuclphysb.2007.06.028
[arXiv:hep-th/0606187 [hep-th]].

%\cite{Lebedev:2006kn}
\bibitem{Lebedev:2006kn}
O.~Lebedev, H.~P.~Nilles, S.~Raby, S.~Ramos-Sanchez, M.~Ratz, P.~K.~S.~Vaudrevange and A.~Wingerter,
%``A Mini-landscape of exact MSSM spectra in heterotic orbifolds,''
Phys. Lett. B \textbf{645}, 88-94 (2007)
%doi:10.1016/j.physletb.2006.12.012
[arXiv:hep-th/0611095 [hep-th]].


%\cite{Ko:2007dz}
\bibitem{Ko:2007dz}
P.~Ko, T.~Kobayashi, J.~h.~Park and S.~Raby,
%``String-derived D(4) flavor symmetry and phenomenological implications,''
Phys. Rev. D \textbf{76}, 035005 (2007)
[erratum: Phys. Rev. D \textbf{76}, 059901 (2007)]
%doi:10.1103/PhysRevD.76.059901
[arXiv:0704.2807 [hep-ph]].

%\cite{Nibbelink:2009sp}
\bibitem{Nibbelink:2009sp}
S.~Groot Nibbelink, J.~Held, F.~Ruehle, M.~Trapletti and P.~K.~S.~Vaudrevange,
%``Heterotic Z(6-II) MSSM Orbifolds in Blowup,''
JHEP \textbf{03}, 005 (2009)
%doi:10.1088/1126-6708/2009/03/005
[arXiv:0901.3059 [hep-th]].



\end{thebibliography}
\end{document}